\documentclass[draftclsnofoot, onecolumn]{IEEEtran}
\usepackage{url,amsmath,graphicx,3parttable,cite}
\usepackage[psamsfonts]{amssymb}
\usepackage[OT2,T1]{fontenc}
\usepackage[russian,english]{babel}
\usepackage[colorlinks, bookmarks=false]{hyperref}

\newcommand{\tr}{\operatorname{tr}}
\newcommand{\var}{\operatorname{var}}

\newtheorem{theorem}{Theorem}
\newtheorem{lemma}{Lemma}
\newtheorem{proposition}{Proposition}
\newtheorem{corollary}{Corollary}

\begin{document}
\title{Statistical Properties of Eigen-Modes and Instantaneous Mutual
Information in MIMO Time-Varying Rayleigh Channels}
\author{\authorblockN{Shuangquan Wang, \IEEEmembership{Student
Member, IEEE, } and Ali Abdi, \IEEEmembership{Member, IEEE}}%
\thanks{This paper was presented in part at the
\emph{40th Ann. Conf. Info. Sci. Sys.}, Princeton, NJ, 2006.}
\thanks{The authors are with the Center for Wireless Communications and Signal
Processing Research (CWCSPR), Department of Electrical and Computer
Engineering, New Jersey Institute of Technology, Newark, NJ 07102
USA (e-mail:\{sw27, ali.abdi\}@njit.edu).}}

\def\citepunct{][}
\def\citedash{]--[}
\maketitle
\begin{abstract}
In this paper, we study two important metrics in multiple-input
multiple-output (MIMO) time-varying Rayleigh flat fading channels.
One is the eigen-mode, and the other is the instantaneous mutual
information (IMI). Their second-order statistics, such as the
correlation coefficient, level crossing rate (LCR), and average
fade/outage duration, are investigated, assuming a general
\emph{nonisotropic} scattering environment. Exact closed-form
expressions are derived and Monte Carlo simulations are provided to
verify the accuracy of the analytical results. For the eigen-modes,
we found they tend to be spatio-temporally uncorrelated in large
MIMO systems. For the IMI, the results show that its correlation
coefficient can be well approximated by the squared amplitude of the
correlation coefficient of the channel, under certain conditions.
Moreover, we also found the LCR of IMI is much more sensitive to the
scattering environment than that of each eigen-mode.
\end{abstract}
\begin{keywords}
Eigen-Modes, Instantaneous Mutual Information, Autocorrelation
Function, Correlation Coefficient, Level Crossing Rate, Average
Fade/Outage Duration, and Multiple-Input Multiple-Output (MIMO).
\end{keywords}

\section{Introduction}\label{sec:introduction}
\PARstart{T}{he} utilization of antenna arrays at the base station
(BS) and the mobile station (MS) in a wireless communication system
increases the capacity linearly with $\min(N_{\!T}, N_{\!R})$, under
certain conditions, where $N_{\!T}$ and $N_{\!R}$ are numbers of
transmit and receive antenna elements, respectively, provided that
the environment is sufficiently rich in multi-path components
\cite{IEEE_sw27:Foschini98, IEEE_sw27:Telatar99_MIMO_Cap}. This is
due to the fact that a multiple-input multiple-output (MIMO) channel
can be decomposed to several parallel single-input single-output
(SISO) channels, called \emph{eigen}-channels or
\emph{eigen}-modes\footnote{We use the two terms interchangeably in
this paper.}, via singular value decomposition
(SVD)\cite{IEEE_sw27:Telatar99_MIMO_Cap,IEEE_sw27:Bliss04_MIMO,
IEEE_sw27:Ivrlac04,IEEE_sw27:Maurer03,IEEE_sw27:Lebrun05_SVD,
IEEE_sw27:Sampath01_SVD,IEEE_sw27:Jayaweera03_MIMO_Cap,
IEEE_sw27:TseBook}.

For a SISO channel, or any subchannel\footnote{In this paper, each
subchannel represents the radio link between each transmit/receive
pair of antennas.} of a MIMO system, there are numerous studies on
key second-order statistics such as correlation, level crossing rate
(LCR), and average fade duration
(AFD)\cite{IEEE_sw27:JakesBook94,IEEE_sw27:Abdi00,
IEEE_sw27:Youssef96,IEEE_sw27:Abdi02_AoA,IEEE_sw27:Abdi03_satellite_model}.
However, to the best of our knowledge, no such study on the
\emph{eigen}-channels of a MIMO system is reported in the
literature, possibly due to the lack of knowledge regarding the
joint probability density function (PDF) of \emph{eigen}-channels.

Regarding another important quantity, the \emph{instantaneous mutual
information} (IMI), only some first-order statistics such as the
mean, variance, outage probability and PDF are
studied\cite{IEEE_sw27:Ozarow94,IEEE_sw27:Hochwald04_Rate_Scheduling,
IEEE_sw27:WangZD04_MIMO_IMI,IEEE_sw27:Smith04_MIMO_IMI_Approx,
IEEE_sw27:TseBook}. Clearly, those statistics do not show the
dynamic temporal behavior, such as correlations, LCR and average
outage durations (AOD) of the IMI in time-varying fading channels.
It is known that IMI can be feedbacked to the rate scheduler in
multi-user communication environments, to increase the system
throughput\cite{IEEE_sw27:Hochwald04_Rate_Scheduling}, where only
the perfect feedback is considered. However, it is hard to obtain
perfect feedback in practice due to the time-varying nature of the
channel, which makes the feedbacked IMI outdated. In this case, the
temporal correlation of IMI can be used to analyze the scheduling
performance with outdated IMI feedbacks. Furthermore, one can
improve the rate scheduling algorithm by exploring the temporal
correlation of IMI.

Several second-order statistics such as the correlation coefficient,
LCR and AOD of IMI in single-input single-output (SISO) systems are
reported in \cite{IEEE_sw27:Wang05_LCR_AFD_SISO_Capacity} and
\cite{IEEE_sw27:Wang05_IT}. For MRC-like MIMO systems, they are
investigated in \cite{IEEE_sw27:Wang05_IT}. However, there are a
limited number of results for a general MIMO channel. In
\cite{IEEE_sw27:Giorgetti03}, some simulation results regarding the
correlation coefficient, LCR and AOD are reported, without
analytical derivations. In \cite{IEEE_sw27:NanZhang05}, lower and
upper bounds, as well as some approximations for the correlation
coefficient of IMI are derived, without exact results at high SNR. A
large gap between the lower and upper bounds and large approximation
errors are observed in \cite[Figs. 2, 5]{IEEE_sw27:NanZhang05}.

In this paper, we extend the results of \cite{IEEE_sw27:Wang05_IT}
to the general MIMO case, using the joint PDF of the
eigenvalues\cite{IEEE_sw27:Wang_SIAM}. Specifically, a number of
second-order statistics such as the autocorrelation function (ACF),
the correlation coefficient, LCR and AFD/AOD\footnote{Note that AFD
is used for eigen-channels, whereas AOD is used for MIMO IMI.} of
the \emph{eigen}-channels and the IMI are studied in MIMO
time-varying Rayleigh flat fading channels. We assume all the
subchannels are spatially independent and identically distributed
(i.i.d.), with the same temporal correlation coefficient,
considering general \emph{nonisotropic} scattering propagation
environments. Closed-form expressions are derived, and Monte Carlo
simulations are provided to verify the accuracy of our closed-form
expressions. The simulation and analytical results show that the
eigen-modes tend to be spatio-temporally uncorrelated in large MIMO
systems, and the correlation coefficient of the IMI can be well
approximated by the squared amplitude of the correlation coefficient
of the channel if $|N_{\!T}-N_{\!R}|$ is much larger than
$\min\left(N_{\!T}, N_{\!R}\right)$. In addition, we also observed
that the LCR of IMI is much more sensitive to the scattering
environment than that of each eigen-mode.

The rest of this paper is organized as follows. Section
\ref{sec:MIMO_Channel} introduces the channel model, as well as the
angle-of-arrival (AoA) model. \emph{Eigen}-channels of a MIMO system
are discussed in Section \ref{sec:eigen-channel}, where Subsection
\ref{ssec:NACF_Coeff_Eigen-Channel} is devoted to the derivation of
the normalized ACF (NACF) and the correlation coefficient of
\emph{eigen}-channels of a MIMO system, whereas Subsection
\ref{ssec:LCR_AFD_Eigen-Channel} focuses on the LCR and AFD of the
\emph{eigen}-channels. The MIMO IMI is investigated in Section
\ref{sec:MIMO-IMI}, in which Subsection
\ref{ssec:ACF_Coeff_MIMO_IMI} addresses the NACF and the correlation
coefficient of the MIMO IMI as well as their low- and high-SNR
approximations, whereas Subsection \ref{ssec:LCR_AOD_MIMO_IMI}
studies the LCR and AOD of the MIMO IMI using the well-known
Gaussian approximation. Numerical results and discussions are
presented in Section \ref{sec:Numerical_Results_Discussion}, and
concluding remarks are given in Section \ref{sec:conclusion}.

{\it Notation}: $\cdot^{\dag}$ is reserved for matrix Hermitian,
$\cdot^{\star}$ for complex conjugate, $\jmath$ for $\sqrt{-1}$,
$\mathbb{E}[\cdot]$ for mathematical expectation, $\mathbf{I}_m$ for
the $m\times m$ identity matrix, $\|\cdot\|_F$ for the Frobenius
norm, $\Re[\cdot]$ and $\Im[\cdot]$ for the real and imaginary parts
of a complex number, respectively, and $f^2(x)$ for
$\left[f(x)\right]^2$. Finally, $t\!\!\in\!\![m,n]$ implies that
$t$, $m$ and $n$ are integers such that $m\leq t\leq n$ with $m\leq
n$.
\section{Channel Model}\label{sec:MIMO_Channel}
In this paper, an $N_{\!R}\times N_{\!T}$ MIMO time-varying Rayleigh
flat fading channel is considered. Similar to
\cite{IEEE_sw27:Ozarow94}, we consider a piecewise constant
approximation for the continuous-time MIMO fading channel matrix
coefficient $\mathbf{H}(t)$, represented by
$\left\{\mathbf{H}(lT_{\!s})\right\}_{l=1}^L$, where $T_{\!s}$ is
the symbol duration and $L$ is the number of samples. In the sequel,
we drop $T_{\!s}$ to simplify the notation. In the $l^\mathrm{th}$
symbol duration, the matrix of the channel coefficients is given by
\begin{equation}\label{eq:H(t)}
\mathbf{H}(l)=\begin{bmatrix}
   h_{1,1}(l) &\cdots  & h_{1,N_{\!T}}(l) \\
   \vdots & \ddots & \vdots \\
   h_{N_{\!R},1}(l) & \cdots & h_{N_{\!R},N_{\!T}}(l) \\
 \end{bmatrix}, l\in[1,L].
\end{equation}

We assume all the $N_{\!T}N_{\!R}$ subchannels
$\left\{h_{n_r,n_t}(l),
l\in[1,L]\right\}_{(n_r=1,n_t=1)}^{(N_{\!R},N_{\!T})}$ are i.i.d.,
with the same temporal correlation coefficient, i.e.,
\begin{equation}\label{eq:Channel_Corr_Assumption}
\mathbb{E}[h_{mn}(l)h_{pq}^\star(l-i)]=\delta_{m,p}\delta_{n,q}
\rho_h(i),
\end{equation}
where the Kronecker delta $\delta_{m,p}$ is $1$ or $0$ when $m=p$ or
$m\neq p$, respectively, and $\rho_h(i)$ is defined and derived at
the end of this section, eq. (\ref{eq:Coeff_h}).

In flat Rayleigh fading channels, each $h_{n_r,n_t}(l), l\in[1,L]$,
is a zero-mean complex Gaussian random process. In the
$l^\mathrm{th}$ interval, $h_{n_r,n_t}(l)$ can be represented
as\cite{IEEE_sw27:Abdi02_AoA}
\begin{equation}\label{eq:Rayleigh_fading_channel}
\begin{split}
h_{n_r,n_t}(l)&=h_{n_r,n_t}^I(l)+\jmath
h_{n_r,n_t}^Q(l),\\&=\alpha_{n_r,n_t}(l)
\exp[-\jmath\Phi_{n_r,n_t}(l)],
\end{split}
\end{equation} where the zero-mean real Gaussian random processes
$h_{n_r,n_t}^I(l)$ and $h_{n_r,n_t}^Q(l)$ are the real and imaginary
parts of $h_{n_r,n_t}(l)$, respectively. $\alpha_{n_r,n_t}(l)$ is
the envelope of $h_{n_r,n_t}(l)$ and $\Phi_{n_r,n_t}(l)$ is the
phase of $h_{n_r,n_t}(l)$. For each $l$, $\alpha_{n_r,n_t}(l)$ has a
Rayleigh distribution and $\Phi_{n_r,n_t}(l)$ is distributed
uniformly over $[-\pi, \pi)$. Without loss of generality, we assume
each subchannel has unit power, i.e.,
$\mathbb{E}[\alpha^2_{n_r,n_t}(l)]=1$.

Using empirically-verified\cite{IEEE_sw27:Abdi02_AoA} multiple von
Mises PDF's\cite[(4)]{IEEE_sw27:Wang05_LCR_AFD_SISO_Capacity} for
the AoA at the receiver in \emph{nonisotropic} scattering
environments, shown as Fig. 1 of
\cite{IEEE_sw27:Wang05_LCR_AFD_SISO_Capacity}, the channel
correlation coefficient of $h_{n_r,n_t}(l)$, $\forall n_r, n_t$, is
given by\cite[(7)]{IEEE_sw27:Wang05_LCR_AFD_SISO_Capacity}
\begin{equation}\label{eq:Coeff_h}
\begin{split}
\rho_h(i)&=\mathbb{E}[h_{n_r,n_t}(l)h_{n_r,n_t}^\star(l-i)],\\
&=\!\sum_{n=1}^N\!P_n\!
\frac{I_0\!\!\left(\!\!\sqrt{\kappa_n^2-4\pi^2f_{\!D}^2i^2T_{\!s}^2
+\jmath4\pi\kappa_nf_{\!D}iT_{\!s}\cos\theta_n}\right)}{I_0(\kappa_n)},
\end{split}
\end{equation} where $I_k(z)=\frac{1}{\pi}\int_0^\pi e^{z\cos w}
\cos(kw)\text{d}w$ is the $k^\text{th}$ order modified Bessel
function of the first kind, $\theta_n$ is the mean AoA of the
$n^{\text{th}}$ cluster of scatterers, $\kappa_n$ controls the width
of the $n^{\text{th}}$ cluster of scatterers, $P_n$ represents the
contribution of the $n^{\text{th}}$ cluster of scatterers such that
$\sum_{n=1}^KP_n=1, 0<P_n \leq 1$, $K$ is the number of clusters of
scatterers, and $f_{\!D}$ is the maximum Doppler frequency. When
$\kappa_n=0, \forall n$, which corresponds to \emph{isotropic}
scattering, (\ref{eq:Coeff_h}) reduces to $\rho_h(i)=I_0(\jmath2\pi
f_{\!D}iT_{\!s})=J_0(2\pi f_{\!D}iT_{\!s})$, which is the Clarke's
correlation model.

\section{Eigen-Channels in MIMO Systems}\label{sec:eigen-channel}
We set $M=\min\left(N_{\!T}, N_{\!R}\right)$ and
$N=\max\left(N_{\!T}, N_{\!R}\right)$. Based on singular value
decomposition
(SVD)\cite{IEEE_sw27:Telatar99_MIMO_Cap,IEEE_sw27:Bliss04_MIMO,
IEEE_sw27:Ivrlac04,IEEE_sw27:Maurer03,IEEE_sw27:Lebrun05_SVD,
IEEE_sw27:Sampath01_SVD,IEEE_sw27:Jayaweera03_MIMO_Cap,
IEEE_sw27:TseBook}, $\mathbf{H}(l)$ in (\ref{eq:H(t)}) can be
diagonalized in the following form
\begin{equation}\label{eq:H(t)_SVD}
\mathbf{H}(l)=\mathbf{U}(l)\mathbf{S}(l)\mathbf{V}^\dag(l),
\end{equation} where $\mathbf{V}(l)$,
whose dimension is $N_{\!T}\times M$, satisfies
$\mathbf{V}^\dag(l)\mathbf{V}(l)=\mathbf{I}_M$, $\mathbf{U}(l)$,
which is $N_{\!R}\times M$, satisfies
$\mathbf{U}^\dag(l)\mathbf{U}(l)=\mathbf{I}_M$, and $\mathbf{S}(l)$
is a diagonal matrix, given by
$\mathbf{S}(l)=\mathrm{diag}\left[s_1(l), \cdots, s_M(l)\right]$, in
which $s_m(l)$, $m\in[1, M]$ is the $m^\mathrm{th}$ non-zero
singular value of $\mathbf{H}(l)$.

\begin{figure}[htbp]
\centering
\includegraphics[width=0.65\linewidth]
{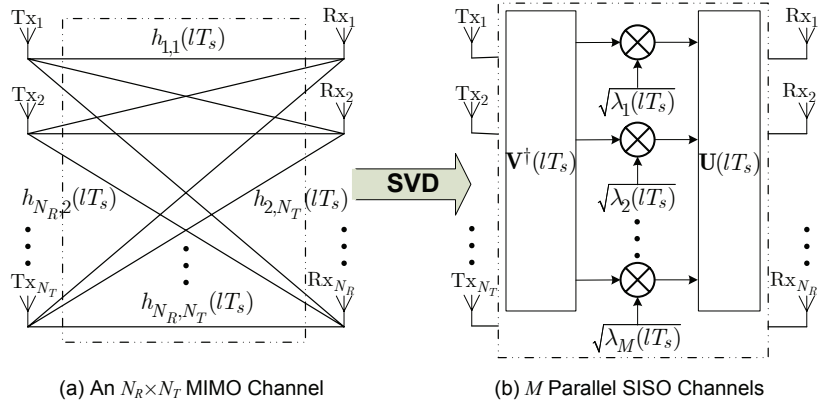} \caption{($a$) A MIMO channel with $N_{\!T}$
transmit and $N_{\!R}$ receive antennas; ($b$) The equivalent $M$
parallel SISO channel representation.}\label{fig:MIMO2SISO}
\end{figure}

We define $\lambda_m(l)=s_m^2(l)$, $\forall m$. Therefore
$\lambda_m(l)$ is the $m^\mathrm{th}$ non-zero eigenvalue of
$\mathbf{H}(l)\mathbf{H}^\dag(l)$. We further consider
$\left\{\lambda_m(l)\right\}_{m=1}^M$ as unordered non-zero
eigenvalues of $\mathbf{H}(l)\mathbf{H}^\dag(l)$. Therefore, the
MIMO channel $\mathbf{H}(l)$ is decomposed to $M$ identically
distributed \emph{eigen}-channels, $\left\{\lambda_m(l),
l\in[1,L]\right\}_{m=1}^M$, by SVD, as shown in Fig.
\ref{fig:MIMO2SISO}. For $M=1$, there is only one
\emph{eigen}-channel, which corresponds to the maximal ratio
transmitter (MRT) if $N_{\!R}=1$, or the maximal ratio combiner
(MRC) if $N_{\!T}=1$. In each case, we have $N$ i.i.d complex
Gaussian branches.

Since all the \emph{eigen}-channels have identical statistics, we
only study one of them and denote it as $\lambda(l), l\in[1,L]$. To
simplify the notation, we use $X$ and $Y$ to denote $\lambda(l)$ and
$\lambda(l-i)$, respectively. The joint PDF of $X$ and $Y$ is given
in (\ref{eq:jpdf_beta_alpha}) of \cite{IEEE_sw27:Wang_SIAM},
\begin{align}
p(x, y)&=\frac{(xy)^\frac{\nu}{2}
e^{-\frac{x+y}{1-\varrho_i^2}}I_\nu\left(\frac{2\varrho_i\sqrt{xy}}
{1-\varrho_i^2}\right)}{M^2(1-\varrho_i^2)\varrho_i^\nu}
\sum_{k=0}^{M-1}\frac{k!L_k^\nu(x)L_k^\nu(y)}
{(k+\nu)!\varrho_i^{2k}}+\frac{(xy)^\nu e^{-(x+y)}}{M^2}\sum_{0\leq
k<l}^{M-1}\bigg\{\frac{k!l!}{(k+\nu)!(l+\nu)!}\nonumber\\
&\hspace{1em}\times\left\{\left[L_k^\nu(x)L_l^\nu(y)\right]^2
+\left[L_l^\nu(x)L_k^\nu(y)\right]^2
-\left[\varrho_i^{2(l-k)}+\varrho_i^{2(k-l)}\right]
L_k^\nu(x)L_l^\nu(x)L_k^\nu(y)L_l^\nu(y)\right\}\!\!\bigg\},
\label{eq:jpdf_beta_alpha}
\end{align} where
$L_n^\alpha(x)=\frac{1}{n!}e^xx^{-\alpha}\frac{\mathrm{d}^n}
{\mathrm{d}x^n}\left(e^{-x}x^{n+\alpha}\right)$ is the associated
Laguerre polynomial of order $n$\cite[pp. 1061,
8.970.1]{IEEE_sw27:RyzhikBook_5th}, $\nu=N-M$, and
$\varrho_i=\left|\rho_h(i)\right|$, where $\rho_h(i)$ is given in
(\ref{eq:Coeff_h}). The joint PDF in (\ref{eq:jpdf_beta_alpha}) is
very general and includes many existing PDF's as special
cases\cite{IEEE_sw27:Wang_SIAM}.
\begin{itemize}
\item By integration over $y$, (\ref{eq:jpdf_beta_alpha}) reduces to the
marginal PDF
\begin{equation}\label{eq:PDF_x}
p(x)=\frac{1}{M}\sum_{m=0}^{M-1}\frac{m!}{(m+\nu)!}
\left[L_m^\nu(x)\right]^2x^\nu e^{-x},
\end{equation} which is the same as the PDF presented
in \cite{IEEE_sw27:Telatar99_MIMO_Cap}. When $M=1$, (\ref{eq:PDF_x})
further reduces to
\begin{equation}\label{eq:PDF_x_M=1}
p(x)=\frac{1}{(N-1)!}x^{N-1} e^{-x},
\end{equation} which is the $\chi^2$ distribution
with $2N$ degrees of freedom\cite[(2.32)]{IEEE_sw27:SimonBook02},
used for characterizing the PDF of outputs of MRT or
MRC\cite{IEEE_sw27:SimonBook04}.

\item With $M=1$, (\ref{eq:jpdf_beta_alpha}) reduces to
\begin{equation}\label{eq:jpdf_x_y_MRC}
p(x,y)=\frac{(xy)^\frac{N-1}{2}\exp\left(\!-\frac{x+y}{1-\varrho_i^2}\!\right)
I_{N-1}\!\left(\frac{2\varrho_i\sqrt{xy}}{1-\varrho_i^2}\right)}
{(N-1)!\left(1\!-\!\varrho_i^2\right)\varrho_i^{N-1}},
\end{equation} which is the joint PDF of outputs of MRT or
MRC at the $l^\mathrm{th}$ and $(l-i)^\mathrm{th}$ symbol
durations\cite{IEEE_sw27:Wang06_Corr_Coeff_MRC}. It includes (3.14)
of \cite{IEEE_sw27:SimonBook02} as a special case\footnote{Eq.
(3.14) in \cite{IEEE_sw27:SimonBook02} is developed for real
uncorrelated Gaussian random variables.}. Furthermore, when $N=1$,
i.e., a SISO channel, (\ref{eq:jpdf_x_y_MRC}) simplifies to
\begin{equation}\label{eq:jpdf_x_y_SISO}
p(x, y)=\frac{1}{1\!-\!\varrho_i^2}\exp\left(\!-\frac{x\!+\!y}
{1\!-\!\varrho_i^2}\!\right)
I_0\!\left(\frac{2\varrho_i\sqrt{xy}}{1\!-\!\varrho_i^2}\right),
\end{equation} which is identical to
(8-103)\cite[pp. 163]{IEEE_sw27:DavenportBook87}, after a one-to-one
nonlinear mapping.
\end{itemize}

In the following subsections, we study the normalized correlation
and correlation coefficient of any two \emph{eigen}-channels,
defined by, respectively,
\begin{equation}\label{eq:NACF_Eigen-Channel_Def}
\widetilde{r}_{m,n}(i)=\frac{\mathbb{E}\left[\lambda_m(l)
\lambda_n(l-i)\right]} {\sqrt{\mathbb{E}\left[\lambda_m^2(l)\right]}
\sqrt{\mathbb{E}\left[\lambda_n^2(l-i)\right]}},
\end{equation} and
\begin{equation}\label{eq:Coeff_Eigen-Channel_Def}
\rho_{m,n}(i)=\frac{\mathbb{E}\left[\lambda_m(l)
\lambda_n(l-i)\right]-\mathbb{E}\left[\lambda_m(l)\right]
\mathbb{E}\left[\lambda_n(l-i)\right]}
{\sqrt{\mathbb{E}\left[\lambda_m^2(l)\right]
-\left\{\mathbb{E}\left[\lambda_m(l)\right]\right\}^2}
\sqrt{\mathbb{E}\left[\lambda_n^2(l-i)\right]
-\left\{\mathbb{E}\left[\lambda_n(l-i)\right]\right\}^2}},
\end{equation}
\subsection{Normalized Correlation and Correlation Coefficient of
Eigen-Channels} \label{ssec:NACF_Coeff_Eigen-Channel} To derive the
normalized correlation and correlation coefficient between any two
eigen-channels, we need the following lemmas.

\begin{lemma}\label{lem:Moments_Eigen-Channel}
The first and second moments of the $m^\mathrm{th}$ eigen-channel
are respectively given by
\begin{align}
\mathbb{E}[\lambda_m(l)]&=N,\label{eq:1stMoment_Eigen-Channel}\\
\mathbb{E}[\lambda^2_m(l)]&=N(N+M).\label{eq:2ndMoment_Eigen-Channel}
\end{align}
\end{lemma}
\begin{proof}
See Appendix \ref{app:Proof_Moments_Eigen-Channel}.
\end{proof}

\begin{lemma}\label{lem:ACF_Eigen-Channel}
The autocorrelation of the $m^\mathrm{th}$ eigen-channel, defined as
$r_{m,m}(i)=\mathbb{E}\left[\lambda_m(l) \lambda_m(l-i)\right]$, is
given by
\begin{equation}\label{eq:ACF_Eigen-Channel}
r_{m,m}(i)=N^2+\frac{N\varrho_i^2}{M}, \quad i\neq0.
\end{equation}
\end{lemma}
\begin{proof}
See Appendix \ref{app:Proof_ACF_Eigen-Channel}
\end{proof}

\begin{lemma}\label{lem:CCF_Eigen-Channel}
The cross-correlation between the $m^\mathrm{th}$ and
$n^\mathrm{th}$ eigen-channels, defined as
$r_{m,n}(i)=\mathbb{E}\left[\lambda_m(l) \lambda_n(l-i)\right]$, is
given by
\begin{equation}\label{eq:CCF_Eigen-Channel}
r_{m,n}(i)=\begin{cases}N^2-N,&i=0,\\
N^2+\frac{N\varrho_i^2}{M}, &i\neq0,
\end{cases}n\neq m.
\end{equation}
\end{lemma}
\begin{proof}
See Appendix \ref{app:Proof_CCF_Eigen-Channel}.
\end{proof}

Based on Lemmas
\ref{lem:Moments_Eigen-Channel}-\ref{lem:CCF_Eigen-Channel}, we
obtain the closed-form expressions for
(\ref{eq:NACF_Eigen-Channel_Def}) and
(\ref{eq:Coeff_Eigen-Channel_Def}), which are given in the following
theorem.
\begin{theorem}
The normalized cross-correlation and the correlation coefficient
between $m^\mathrm{th}$ and $n^\mathrm{th}$ eigen-channels, defined
in (\ref{eq:NACF_Eigen-Channel_Def}) and
(\ref{eq:Coeff_Eigen-Channel_Def}), are respectively given by
\begin{equation}\label{eq:NACF_Eigen-Channel}
\widetilde{r}_{m,n}(i)=
\begin{cases} \frac{M-(M+1)\left(1-\delta_{m,n}\right)}{N+M},&i=0,\\
\frac{M\!N+\varrho_i^2}{M\!N+M^2}, &i\neq0,
\end{cases}
\end{equation} and
\begin{equation}\label{eq:Coeff_Eigen-Channel}
\rho_{m,n}(i)=\begin{cases} 1-\frac{M+1}{M}\left(1-\delta_{m,n}\right),&i=0,\\
\frac{\varrho_i^2}{M^2}, &i\neq0.
\end{cases}
\end{equation}
\end{theorem}
\begin{proof}
From Lemma \ref{lem:Moments_Eigen-Channel}, it is straightforward to
see that the eigen-channel is stationary in the wide sense.
Moreover, all the eigen-channels have the same statistics, therefore
we have
$\mathbb{E}\left[\lambda_n(l-i)\right]=\mathbb{E}\left[\lambda_m(l)\right]$
and
$\mathbb{E}\left[\lambda_n^2(l-i)\right]=\mathbb{E}\left[\lambda_m^2(l)\right]$,
$\forall m,n\in[1,M]$ and $\forall l,i$. By plugging
(\ref{eq:2ndMoment_Eigen-Channel})-(\ref{eq:ACF_Eigen-Channel}) into
(\ref{eq:NACF_Eigen-Channel_Def}), we obtain
(\ref{eq:NACF_Eigen-Channel}). Finally, substitution of
(\ref{eq:1stMoment_Eigen-Channel})-(\ref{eq:ACF_Eigen-Channel}) into
(\ref{eq:Coeff_Eigen-Channel_Def}) results in
(\ref{eq:Coeff_Eigen-Channel}).
\end{proof}

From (\ref{eq:NACF_Eigen-Channel}) and
(\ref{eq:Coeff_Eigen-Channel}), we have the following interesting
observations.
\begin{itemize}
\item If $M$ is greater than $1$, the normalized correlation and the
correlation coefficient are not continuous at $i=0$, as
$\tilde{r}_{m,n}(1)$ and $\rho_{m,n}(1)$ do not converge to
$\tilde{r}_{m,n}(0)=\rho_{m,n}(0)=1$ as $T_{\!s}\rightarrow0$,
$\forall m,n$.
\item If $M$ is large, all the $M$ eigen-channels tend to be
spatio-temporally uncorrelated, due to
\begin{equation}\label{eq:Coeff_Eigen-Channel_Asymtotic}
\lim_{M\rightarrow\infty}\rho_{m,n}(i)=\delta_{m,n}\delta_{i,0}.
\end{equation}
\end{itemize}

As an example, with isotropic scattering,
(\ref{eq:NACF_Eigen-Channel}) and (\ref{eq:Coeff_Eigen-Channel}),
respectively, reduce to
\begin{equation}\label{eq:NACF_Eigen-Channel_Iso}
\widetilde{r}_{m,n}(i)=
\begin{cases} \frac{M-(M+1)\left(1-\delta_{m,n}\right)}{N+M},&i=0,\\
\frac{M\!N+J_0^2\left(2\pi f_{\!D}iT_{\!s}\right)}{M\!N+M^2},
&i\neq0,
\end{cases}
\end{equation} and
\begin{equation}\label{eq:Coeff_Eigen-Channel_Iso}
\rho_{m,n}(i)=\begin{cases} 1-\frac{M+1}{M}\left(1-\delta_{m,n}\right),&i=0,\\
\frac{J_0^2\left(2\pi f_{\!D}iT_{\!s}\right)}{M^2}, &i\neq0,
\end{cases}
\end{equation}

\subsection{LCR and AFD of an Eigen-Channel}
\label{ssec:LCR_AFD_Eigen-Channel} In this subsection, we calculate
the LCR and AFD of an eigen-channel at a given level. To simplify
the notation, the eigen-channel index $m$ is dropped in this
subsection, as the derived LCR and AFD results hold for any
eigen-channel.
\subsubsection{LCR of an Eigen-Channel}\label{sssec:LCR_Eigen-Channel}
Similar to the calculation of zero crossing rate in discrete
time\cite[Ch. 4]{IEEE_sw27:KedemBook94}, we define the binary
sequence $\left\{Z_l\right\}_{l=1}^L$, based on the
\emph{eigen}-channel samples $\left\{\lambda(l)\right\}_{l=1}^L$, as
\begin{equation}\label{eq:Hardening_Function}
Z_l=\begin{cases} 1,&\text{if } \lambda(l)\geq
\lambda_\text{th},\\
0,&\text{if } \lambda(l)<\lambda_\text{th},
\end{cases}
\end{equation} where $\lambda_\text{th}$ is a fixed threshold. The
number of crossings of $\left\{\lambda(l)\right\}_{l=1}^L$ with
$\lambda_\text{th}$, within the time interval $T_{\!s}\leq t\leq
LT_{\!s}$, denoted by $D_{\lambda_\text{th}}$, can be defined in
terms of $\{Z_l\}_{l=1}^L$\cite[(4.1)]{IEEE_sw27:KedemBook94}
\begin{equation}\label{eq:Def_Num_UpDownCrossings}
D_{\lambda_\text{th}}=\sum_{l=2}^L\left(Z_l-Z_{l-1}\right)^2,
\end{equation} which includes both up- and down-crossings.

After some simple manipulations, the expected crossing rate at the
level $\lambda_\text{th}$ can be written as
\begin{equation}\label{eq:Def_LCR_UpDown_DT}
\frac{\mathbb{E}[D_{\lambda_\text{th}}]}{(L-1)T_{\!s}}
=\frac{2P_r\{Z_l=1\}-2P_r\{Z_l=1,Z_{l-1}=1\}}{T_{\!s}},
\end{equation} where $P_r\{\cdot\}$ is the probability of an event.
Therefore, the expected down crossing rate at
$\lambda_\text{th}$, denoted by $N_\lambda(\lambda_\text{th})$, is
half of (\ref{eq:Def_LCR_UpDown_DT}), given by
\begin{equation}\label{eq:Def_LCR_Up_DT}
N_\lambda(\lambda_\text{th})=\frac{\phi_\lambda(\lambda_\mathrm{th})
-\varphi_\lambda(\lambda_\mathrm{th})}{T_{\!s}},
\end{equation} where $\phi_\lambda(\lambda_\mathrm{th})=P_r\{Z_l=1\}$
and $\varphi_\lambda(\lambda_\mathrm{th})=P_r\{Z_l=1,Z_{l-1}=1\}$.
Analytical expressions for $\phi_\lambda(\lambda_\mathrm{th})$ and
$\varphi_\lambda(\lambda_\mathrm{th})$ are stated in the following
theorem.

\begin{theorem}\label{theo:phi_varphi_rho1}
For a given threshold $\lambda_\text{th}$,
$\phi_\lambda(\lambda_\mathrm{th})$ and
$\varphi_\lambda(\lambda_\mathrm{th})$ are, respectively, given by
\begin{equation}\label{eq:phi_Eigen}
\phi_\lambda(\lambda_\mathrm{th})\!=\!\frac{1}{M}\!\sum_{m=0}^{M-1}\!
\sum_{p=0}^m\!\sum_{q=0}^m\!\frac{m!{m+\nu\choose m-p}{m+\nu\choose
m-q}\Gamma(p\!+\!q\!+\!\nu\!+\!1,\lambda_\text{th})}{(m\!+\!\nu)!p!q!(-1)^{p+q}},
\end{equation} and
\begin{equation}\label{eq:varphi_Eigen}
\varphi_\lambda(\lambda_\mathrm{th})=\phi^2_\lambda(\lambda_\mathrm{th})+\frac{1}{M^2}\sum_{j=M}^\infty
\sum_{k=0}^{M-1}\frac{j!k!\varrho_1^{2(j-k)}}{(j+\nu)!(k+\nu)!}\left[\sum_{p=0}^j\sum_{q=0}^k\frac{{j+\nu\choose
j-p}{k+\nu\choose
k-q}}{p!q!(-1)^{p+q}}\Gamma(p+q+\nu+1,\lambda_\text{th})\right]^2,
\end{equation} where $\Gamma(a,z)=\int_z^\infty
t^{a-1}e^{-t}\mathrm{d}t$\cite[pp.~949,
8.350.2]{IEEE_sw27:RyzhikBook_5th} is the upper incomplete gamma
function, ${n\choose k}$ is the binomial coefficient, given by
$\frac{n!}{k!(n-k)!}$, and $\varrho_1=|\rho_h(1)|$, defined before,
i.e.,
\begin{equation}\label{eq:varrho_1}
\varrho_1\!=\!\left|\sum_{n=1}^N\!P_n\!
\frac{I_0\!\!\left(\!\sqrt{\kappa_n^2\!-\!4\pi^2f_{\!D}^2T_{\!s}^2
\!+\!\jmath4\pi\kappa_nf_{\!D}T_{\!s}\cos\theta_n}\right)}
{I_0(\kappa_n)}\!\right|.
\end{equation}
\end{theorem}
\begin{proof}
$L_n^\nu(x)$ is a polynomial of order $n$, and can be represented
as\cite[pp.~1061, 8.970.1]{IEEE_sw27:RyzhikBook_5th}
\begin{equation}\label{eq:Laguerre_Poly_SeriesFormat}
L_n^\nu(x)=\sum_{k=0}^n{n+\nu\choose n-k}\frac{(-x)^k}{k!}.
\end{equation} By plugging (\ref{eq:Laguerre_Poly_SeriesFormat})
into (\ref{eq:PDF_x}), the univariate PDF of an eigen-channel, and
integrating over $x$ from $\lambda_\text{th}$ to $\infty$, we obtain
(\ref{eq:phi_Eigen}). Similarly, substitution of
(\ref{eq:Laguerre_Poly_SeriesFormat}) into
(\ref{eq:jpdf_beta_alpha_alternate_way}), the bivariate PDF of an
eigen-channel, and integration over $x$ from $\lambda_\text{th}$ to
$\infty$ results in (\ref{eq:varphi_Eigen}).
\end{proof}

By plugging (\ref{eq:phi_Eigen}) and (\ref{eq:varphi_Eigen}) into
(\ref{eq:Def_LCR_Up_DT}), we obtain the expected crossing rate at
the level $\lambda_\text{th}$.
\subsubsection{AFD of an Eigen-Channel}\label{sssec:AFD_Eigen-Channel}
The cumulative distribution function (CDF) of $\lambda(l), \forall
l$, is obtained as
\begin{equation}\label{eq:CDF_Eigen-Channel}
F_\lambda(\lambda_\text{th})=P_r\left\{X\leq
\lambda_\text{th}\right\}=1-\phi_\lambda(\lambda_\mathrm{th}),
\end{equation} where $\phi_\lambda(\lambda_\mathrm{th})$ is given in
(\ref{eq:phi_Eigen}).

The AFD of the \emph{eigen}-channel
$\left\{\lambda(l)\right\}_{l=1}^L$ is therefore given by
\begin{equation}\label{eq:AFD_Eigen-Channel}
\overline{t}_\lambda(\lambda_\text{th})
=\frac{F_\lambda(\lambda_\text{th})}{N_\lambda(\lambda_\text{th})}
=\frac{\left[1-\phi_\lambda(\lambda_\mathrm{th})\right]\!T_{\!s}}
{\phi_\lambda(\lambda_\mathrm{th})-
\varphi_\lambda(\lambda_\mathrm{th})},
\end{equation} where $\phi_\lambda(\lambda_\mathrm{th})$ and
$\varphi_\lambda(\lambda_\mathrm{th})$ are given in
(\ref{eq:phi_Eigen}) and (\ref{eq:varphi_Eigen}), respectively.

\section{MIMO IMI}\label{sec:MIMO-IMI}
In this section, the NACF, the correlation coefficient, LCR and AOD
of IMI in a MIMO system are investigated in detail. In the presence
of the additive white Gaussian noise, if perfect channel state
information $\left\{\mathbf{H}(l)\right\}_{l=1}^L$, is available at
the receiver only, the ergodic channel capacity is given
by\cite{IEEE_sw27:Telatar99_MIMO_Cap,IEEE_sw27:TseBook}
\begin{equation}\label{eq:Cap_MIMO}
C=\mathbb{E}\left[\ln\det\left(\mathbf{I}_{N_{\!R}}+\frac{\eta}{N_{\!T}}
\mathbf{H}_l\mathbf{H}_l^\dag\right)\right],
\end{equation} in nats/s/Hz, where $\eta$ is the average SNR at each
receive antenna, and $\mathbf{H}_l$ denotes $\mathbf{H}(l)$.

In the above equation, at any given time index $l$,
$\ln\det\left(\mathbf{I}_{N_{\!R}}+\frac{\eta}{N_{\!T}}\mathbf{H}_l
\mathbf{H}_l^\dag\right)$ is a random variable as it depends on the
random channel matrix $\mathbf{H}_l$. Therefore
\begin{equation}\label{eq:Def_IMI_MIMO}
\mathcal{I}_l=\ln\det\left(\mathbf{I}_{N_{\!R}}+\frac{\eta}{N_{\!T}}\mathbf{H}_l
\mathbf{H}_l^\dag\right), \quad l=1,2,\cdots,
\end{equation} is a discrete-time random process with
the ergodic capacity as its mean.

By plugging (\ref{eq:H(t)_SVD}) into (\ref{eq:Def_IMI_MIMO}), we can
express the IMI in terms of $M$ eigenvalues as
\begin{equation}\label{eq:MIMO_IMI_Eigenvalue}
\mathcal{I}_l=\sum_{m=1}^M\ln\left(1+\frac{\eta}{N_{\!T}}\lambda_m(l)\right),
\quad l=1,2,\cdots.
\end{equation}

\subsection{NACF and Correlation
Coefficient of MIMO IMI} \label{ssec:ACF_Coeff_MIMO_IMI} In this
subsection, we derive exact closed-form expression fors the NACF and
the correlation coefficient of MIMO IMI, and their approximations at
low- and high-SNR regimes, using the following lemmas.
\begin{lemma}\label{lem:Moments_MIMO_IMI}
The mean and second moment of $\mathcal{I}_l$ are respectively given
by (\ref{eq:1stMoment_MIMO_IMI}) and (\ref{eq:2ndMoment_MIMO_IMI})
\begin{equation}\label{eq:1stMoment_MIMO_IMI}
\mathbb{E}[\mathcal{I}_l]=\sum_{m=0}^{M-1}
\sum_{p=0}^m\sum_{q=0}^m\frac{m!{m+\nu\choose m-p}{m+\nu\choose
m-q}}{(m\!+\!\nu)!p!q!(-1)^{p+q}}G_{2,3}^{3,
1}\!\!\left(\!\frac{N_{\!T}}{\eta}\!\left|\!\!
\begin{array}{c}
0,1 \\
0,0,p+q+\nu+1
\end{array}\right.
\!\!\!\!\right),
\end{equation}
\begin{multline}\label{eq:2ndMoment_MIMO_IMI}
\mathbb{E}[\mathcal{I}_l^2]=
2e^\frac{N_{\!T}}{\eta}\sum_{m=0}^{M-1}\sum_{p=0}^m\sum_{q=0}^m
\sum_{j=0}^{p+q+\nu}\frac{(-1)^{\nu-j}m!{m+\nu\choose
m-p}{m+\nu\choose m-q}{p+q+\nu\choose
j}}{\left(\frac{\eta}{N_{\!T}}\right)^{p+q+\nu+1}
(m\!+\!\nu)!p!q!}G_{3,4}^{4,
0}\!\!\left(\!\frac{N_{\!T}}{\eta}\!\left|\!\!
\begin{array}{c}
-j,-j,-j \\
0,-j-1,-j-1,-j-1
\end{array}\right.
\!\!\!\!\right)
\\-\sum_{j=0}^{M\!-\!1}
\!\sum_{k=0}^{M\!-\!1}\!\frac{j!k!}{(j\!+\!\nu)!(k\!+\!\nu)!}
\!\left[\sum_{p=0}^j\sum_{q=0}^k\frac{{j+\nu\choose
j-p}{k+\nu\choose k-q}}{p!q!(-1)^{p+q}}G_{2,3}^{3,
1}\!\!\left(\!\frac{N_{\!T}}{\eta}\!\left|\!\!
\begin{array}{c}
0,1 \\
0,0,p\!+\!q\!+\!\nu\!+\!1
\end{array}\right.
\!\!\!\!\right)\!\!\right]^2+\{\mathbb{E}[\mathcal{I}_l]\}^2,
\end{multline} where $G$ is Meijer's $G$ function\cite[pp.~1096,
9.301]{IEEE_sw27:RyzhikBook_5th}.
\end{lemma}
\begin{proof}
See Appendix \ref{app:Proof_Moments_MIMO_IMI}.
\end{proof}

\begin{lemma}\label{lem:ACF_MIMO_IMI}
The ACF of MIMO IMI, defined as
$r_{\!\mathcal{I}}(i)=\mathbb{E}[\mathcal{I}_l\mathcal{I}_{l-i}]$,
is shown to be
\begin{equation}\label{eq:ACF_MIMO_IMI}
r_{\!\mathcal{I}}(i)=\left\{\mathbb{E}[\mathcal{I}_l]\right\}^2
+\sum_{j=M}^\infty
\sum_{k=0}^{M-1}\frac{j!k!\varrho_i^{2(j-k)}}{(j\!+\!\nu)!(k\!+\!\nu)!}
\!\left[\sum_{p=0}^j\sum_{q=0}^k\frac{{j+\nu\choose
j-p}{k+\nu\choose k-q}}{p!q!(-1)^{p+q}}G_{2,3}^{3,
1}\!\!\left(\!\frac{N_{\!T}}{\eta}\!\left|\!\!
\begin{array}{c}
0,1 \\
0,0,p\!+\!q\!+\!\nu\!+\!1
\end{array}\right.
\!\!\!\!\right)\!\!\right]^2\!\!\!\!.
\end{equation}
\end{lemma}
\begin{proof} By plugging (\ref{eq:Laguerre_Poly_SeriesFormat})
into (\ref{eq:jpdf_beta_alpha_alternate_way}), and using
(\ref{eq:MeijerG_for_Integrand}), we obtain (\ref{eq:ACF_MIMO_IMI})
immediately.
\end{proof}

With Lemmas \ref{lem:Moments_MIMO_IMI} and \ref{lem:ACF_MIMO_IMI},
the NACF and the correlation coefficient can be calculated according
to
\begin{equation}\label{eq:Def_NACF_MIMO_IMI}
\tilde{r}_{\!\mathcal{I}}(i)=\frac{r_{\!\mathcal{I}}(i)}
{\mathbb{E}[\mathcal{I}_l^2]},
\end{equation}
and
\begin{equation}\label{eq:Def_Coeff_MIMO_IMI}
\rho_{\!\mathcal{I}}(i)=\frac{r_{\!\mathcal{I}}(i)-\{\mathbb{E}[\mathcal{I}_l]\}^2}
{\mathbb{E}[\mathcal{I}_l^2]-\{\mathbb{E}[\mathcal{I}_l]\}^2},
\end{equation} by inserting (\ref{eq:2ndMoment_MIMO_IMI}) and
(\ref{eq:1stMoment_MIMO_IMI}) into (\ref{eq:Def_NACF_MIMO_IMI}), and
(\ref{eq:2ndMoment_MIMO_IMI}), (\ref{eq:1stMoment_MIMO_IMI}) and
(\ref{eq:ACF_MIMO_IMI}) into (\ref{eq:Def_Coeff_MIMO_IMI}),
respectively.

In general, it seems difficult to further simplify
(\ref{eq:2ndMoment_MIMO_IMI}), (\ref{eq:1stMoment_MIMO_IMI}) and
(\ref{eq:ACF_MIMO_IMI}). However, we note that
\begin{equation}\label{eq:Xi_LowHighSNR_MIMO}
\ln(1+\omega x)\approx
\begin{cases}
\omega x,& \omega\rightarrow0,\\
\ln(\omega x),& \omega\rightarrow\infty.
\end{cases}
\end{equation}

Using (\ref{eq:Xi_LowHighSNR_MIMO}), we obtain asymptotic
closed-form expressions for the NACF,
$\tilde{r}_{\!\mathcal{I}}(i)$, and the correlation coefficient,
$\rho_{\!\mathcal{I}}(i)$, at low- and high-SNR regimes, as follows.
\subsubsection{The Low-SNR Regime} \label{sssec:ACF_Coeff_LowSNR_MIMO_IMI} If
$\eta\rightarrow0$, based on (\ref{eq:Xi_LowHighSNR_MIMO}),
(\ref{eq:MIMO_IMI_Eigenvalue}) can be approximated by
\begin{equation}\label{eq:MIMO_IMI_LowSNR_Approx}
\mathcal{I}_l\approx\sum_{m=1}^{M}\frac{\eta}{N_{\!T}} \lambda_m(l),
\end{equation} which is the same as the low-SNR approximation of
$\mathcal{I}_l$ in a MIMO system with orthogonal space-time block
code (OSTBC) transmission\cite{IEEE_sw27:Wang05_IT}, due to
$\sum_{m=1}^{M}\lambda_m(l)=\tr\left[\mathbf{H}_l\mathbf{H}_l^\dag\right]
=\|\mathbf{H}_l\|_F^2$. Therefore, the NACF and correlation
coefficient of interest are equal to those derived for the
OSTBC-MIMO system at low SNRs, as stated in the following
proposition.

\begin{proposition}
At the low-SNR regime, the NACF and the correlation coefficient are
given by\cite{IEEE_sw27:Wang05_IT}
\begin{align}
\tilde{r}_{\!\mathcal{I}}(i)&\approx\frac{N_{\!R}N_{\!T}+\varrho_i^2}
{N_{\!R}N_{\!T}+1},
\label{eq:NACF_LowSNR_MIMO_IMI}\\
\rho_{\!\mathcal{I}}(i)&\approx\varrho_i^2.\label{eq:Coeff_LowSNR_MIMO_IMI}
\end{align}
\end{proposition}

\subsubsection{The High-SNR Regime} \label{sssec:ACF_Coeff_HighSNR_MIMO}
If $\eta\rightarrow\infty$, based on (\ref{eq:Xi_LowHighSNR_MIMO}),
(\ref{eq:MIMO_IMI_Eigenvalue}) can be approximated by
\begin{equation}\label{eq:MIMO_IMI_HighSNR_Approx}
\mathcal{I}_l\approx\sum_{m=1}^{M}\ln\left[\frac{\eta}{N_{\!T}}
\lambda_m(l)\right],
\end{equation} whose NACF and correlation coefficient are
presented in the following theorem.

\begin{theorem}\label{theo:NACF_Coeff_MIMO_IMI_HighSNR_Approx}
At high SNRs, the NACF and the correlation coefficient are given by
(\ref{eq:NACF_HighSNR_MIMO_IMI}) and
(\ref{eq:Coeff_HighSNR_MIMO_IMI}), respectively
\begin{align}
\tilde{r}_{\!\mathcal{I}}(i)&\approx\frac{\sum_{m=0}^{M\!-\!1}\!
\frac{M!(m+\nu)!\varrho_i^{2(M-m)}{}_4F_3(M-m,M-m,M+1,1;M-m+1,
M-m+1,N+1;\varrho_i^2)}{(M-m)^2N!m!}
\!+\!\left(\sum_{m=0}^{M-1}\!\psi_{\!N-m}
+M\ln\frac{\eta}{N_{\!T}}\right)^2}
{\sum_{m=0}^{M-1}\zeta(2,N-m)+\left(\sum_{m=0}^{M-1}\!\psi_{\!N-m}
+M\ln\frac{\eta}{N_{\!T}}\right)^2},
\label{eq:NACF_HighSNR_MIMO_IMI}\\
\rho_{\!\mathcal{I}}(i)&\approx \frac{\sum_{m=0}^{M-1}
\frac{M!(m+\nu)!\varrho_i^{2(M-m)}}{(M-m)^2N!m!}
{}_4F_3(M-m,M-m,M+1,1;M-m+1,M-m+1,N+1;\varrho_i^2)}
{\sum_{m=0}^{M-1}\zeta(2,N-m)},\label{eq:Coeff_HighSNR_MIMO_IMI}
\end{align} where
$_pF_q\left(a_1,\cdots,a_p;b_1,\cdots,b_q;z\right)$ is the
generalized hypergeometric function \cite[pp. 1071,
9.14.1]{IEEE_sw27:RyzhikBook_5th}, $\zeta(\cdot,\cdot)$ is the
Riemann zeta function, given by
$\zeta(z,q)=\sum_{k=0}^\infty\frac{1}{(q+k)^z}$\cite[pp.~1101,
9.521.1]{IEEE_sw27:RyzhikBook_5th}, and $\psi_k$ is the digamma
function\cite[pp. 954, 8.365.4]{IEEE_sw27:RyzhikBook_5th}.
\end{theorem}
\begin{proof}
See Appendix \ref{app:NACF_Coeff_MIMO_IMI_HighSNR_Approx}.
\end{proof}

Theorem \ref{theo:NACF_Coeff_MIMO_IMI_HighSNR_Approx} includes the
high-SNR approximation for the OSTBC-MIMO system in
\cite{IEEE_sw27:Wang05_IT} as a special case. In fact, with $M=1$,
(\ref{eq:NACF_HighSNR_MIMO_IMI}) and
(\ref{eq:Coeff_HighSNR_MIMO_IMI}) simplify to the corresponding
resutls in \cite{IEEE_sw27:Wang05_IT} by replacing $N$ with $M\!N$,
i.e.,
\begin{align}\label{eq:NACF_Coeff_HighSNR_OSTBC}
\tilde{r}_{\!\mathcal{I}}(i)&\!\approx\!\frac{\frac{\varrho_i^2}{M\!N}
{}_3F_2\!\left(\!1,1,1;2,M\!N\!+\!1;\varrho_i^2\!\right)
\!+\!\left(\!\psi_{\!M\!N}\! +\!\ln\frac{\eta}{N_{\!T}}\!\right)^2}
{\zeta(2,M\!N)+\left(\psi_{\!M\!N}
+\ln\frac{\eta}{N_{\!T}}\right)^2},\\
\rho_{\!\mathcal{I}}(i)&\!\approx\!\frac{\frac{\varrho_i^2}{M\!N}
{}_3F_2\left(1,1,1;2,M\!N+1;\varrho_i^2\right)}{\zeta(2,M\!N)},
\end{align} where the identity
${}_4F_3\left(1,1,1,2;2,2,M\!N+1;x\right)
={}_3F_2\left(1,1,1;2,M\!N+1;x\right)$ is used.

Based on Theorem \ref{theo:NACF_Coeff_MIMO_IMI_HighSNR_Approx}, we
conclude that if $\nu=0$ and $M\rightarrow\infty$,
(\ref{eq:Coeff_HighSNR_MIMO_IMI}) reduces to
\begin{equation}\label{eq:Coeff_HighSNR_MIMO_IMI_Asymp}
\lim_{M\rightarrow\infty}\rho_\mathcal{I}(i)
=\frac{-\ln\left(1-\varrho_i^2\right)}
{\lim_{p\rightarrow\infty}\sum_{k=1}^p\frac{1}{k}}=\delta_{i,0},
\end{equation} where we the first ``='' is obtained by collecting
the terms in (\ref{eq:Coeff_HighSNR_MIMO_IMI}), and the second ``=''
is due to $\varrho_i<1, i\neq0$. We conjecture that the second ``=''
of (\ref{eq:Coeff_HighSNR_MIMO_IMI_Asymp}) holds for any finite
$\nu$ at high SNRs, i.e.,
$\lim_{M\rightarrow\infty}\rho_\mathcal{I}(i)=\delta_{i,0}$,
$\forall \nu<\infty$. It implies that MIMO IMI is asymptotically
uncorrelated at high SNRs, if the difference between the numbers of
Tx and Rx antennas is finite.

To better understand Theorem
\ref{theo:NACF_Coeff_MIMO_IMI_HighSNR_Approx}, the Taylor expansion
of (\ref{eq:Coeff_HighSNR_MIMO_IMI}) and the maximum difference
between (\ref{eq:Coeff_LowSNR_MIMO_IMI}) and
(\ref{eq:Coeff_HighSNR_MIMO_IMI}) is listed in Table
\ref{tab:Taylor_rho_I}, for different values of $M$ and $N$. From
Table \ref{tab:Taylor_rho_I}, the following observations can be
made.
\begin{itemize}
\item If $\nu=N-M$ is fixed, the maximum difference
between the low- and high-SNR approximations increases when $M$
increases, which is supported by the first four rows of Table
\ref{tab:Taylor_rho_I}, i.e., $(M,N)=(1,1)$, $(2,2)$, $(3,3)$, and
$(4,4)$.
\item From the last several rows of Table
\ref{tab:Taylor_rho_I}, i.e., $(M,N)=(4,4)$, $(4,8)$, $(4,12)$ and
$(4,16)$, one may conclude that if $M$ is fixed, the maximum
difference between the low- and high-SNR approximations decreases as
$\nu$ increases. Furthermore, $\rho_{\!\mathcal{I}}(i)$ can be well
approximated by $\varrho_i^2$, with negligible error for any SNR,
when $\frac{\nu}{M}$ is not small.
\end{itemize}

\subsection{LCR and AOD of MIMO IMI}
\label{ssec:LCR_AOD_MIMO_IMI} The technique developed in Subsection
\ref{ssec:LCR_AFD_Eigen-Channel} is also valid for calculating the
LCR and AOD of MIMO IMI, i.e., we can obtain them by replacing
$\phi_\lambda(\lambda_\mathrm{th})$ and
$\varphi_\lambda(\lambda_\mathrm{th})$ with
$\phi_{\!\mathcal{I}}(I_\mathrm{th})=P_r\{\mathcal{I}_l>I_\mathrm{th}\}$
and
$\varphi_{\!\mathcal{I}}(I_\mathrm{th})=P_r\{\mathcal{I}_l>I_\mathrm{th},
\mathcal{I}_{l-1}>I_\mathrm{th}\}$ in (\ref{eq:Def_LCR_Up_DT}) and
(\ref{eq:AFD_Eigen-Channel}), respectively. Therefore, we only need
to calculate, $\phi_{\!\mathcal{I}}(I_\mathrm{th})$ and
$\varphi_{\!\mathcal{I}}(I_\mathrm{th})$, which are presented in the
following theorem.
\begin{theorem}\label{theo:phi_varphi_MIMO_IMI} At any given
level $I_\mathrm{th}$, $\phi_{\!\mathcal{I}}(I_\mathrm{th})$ and
$\varphi_{\!\mathcal{I}}(I_\mathrm{th})$ can be expressed in terms
of multiple integrals, given by (\ref{eq:phi_MIMO_IMI}) and
(\ref{eq:varphi_MIMO_IMI}), respectively.
\begin{align}
\phi_{\!\mathcal{I}}(I_\mathrm{th})&=\overbrace{\int\cdots\int}^{M}_{\prod_{m=1}^M
\left(1+\frac{\eta\,x_m}{N_{\!T}}\!\right)>\exp(I_\mathrm{th})}
\underbrace{\frac{\prod_{m=1}^Mx_m^\nu\prod_{m<n}^M(x_m-
x_n)^2}{M!\prod_{m=0}^{M-1}m!(m+\nu)!\exp\left(\sum_{m=1}^Mx_m\right)}}_{p(x_1,
x_2, \cdots, x_{\!M})} \prod_{m=1}^Mdx_m,
\label{eq:phi_MIMO_IMI}\\
\varphi_{\!\mathcal{I}}(I_\mathrm{th})&=
\overbrace{\int\cdots\int}^{2M}_{\begin{subarray}{c}\prod_{m=1}^M
\left(1+\frac{\eta\,x_m}{N_{\!T}}\!\right)>\exp(I_\mathrm{th})
\\\prod_{m=1}^M
\left(1+\frac{\eta\,y_m}{N_{\!T}}\!\right)>\exp(I_\mathrm{th})\end{subarray}}
\underbrace{\frac{\prod_{m<n}^M\!\left[(x_m\!-\!x_n)(y_m\!-\!
y_n)\right]^2\prod_{m=1}^M \!(\sqrt{x_my_m})^\nu \det\left|
I_\nu\!\!\left(\frac{2\varrho_1\sqrt{x_my_n}}{1-\varrho_1^2}\right)\!\right|}
{M!M!\prod_{m=0}^{M-1}m!(m+\nu)!\varrho_1^{M\!N-M}(1-\varrho_1^2)^M\exp
\left(\frac{\sum_{m=1}^Mx_m+y_m} {1-\varrho_1^2}\right)}} _{p(x_1,
x_2, \cdots, x_{\!M}, y_1, y_2, \cdots,
y_{\!M})}\prod_{m=1}^Mdx_mdy_m.\label{eq:varphi_MIMO_IMI}
\end{align}
\end{theorem}
\begin{proof}
Let $\{X_m\}_{m=1}^M$ and $\{Y_m\}_{m=1}^M$ be $M$ unordered
eigenvalues of $\mathbf{H}(l)\mathbf{H}^\dag(l)$ and
$\mathbf{H}(l-1)\mathbf{H}^\dag(l-1)$, respectively. Then the joint
PDF of $\{X_m\}_{m=1}^M$ is given by ${p(x_1, x_2, \cdots,
x_{\!M})}$ in
(\ref{eq:phi_MIMO_IMI})\cite{IEEE_sw27:James64_MatrixVariate}, and
the joint PDF of $\{X_m\}_{m=1}^M$ and $\{Y_m\}_{m=1}^M$ is given by
${p(x_1, x_2, \cdots, x_{\!M}, y_1, y_2, \cdots, y_{\!M})}$ in
(\ref{eq:varphi_MIMO_IMI})\cite{IEEE_sw27:Wang_SIAM}. Moreover,
according to (\ref{eq:MIMO_IMI_Eigenvalue}), the event
$\left\{\mathcal{I}_l>I_\mathrm{th}\right\}$ is equivalent to
$\left\{\prod_{m=1}^M
\left(1+\frac{\eta\,X_m}{N_{\!T}}\!\right)>e^{I_\mathrm{th}}\right\}$,
which leads to (\ref{eq:phi_MIMO_IMI}). Similarly, it is
straightforward to see that the two events
$\left\{\mathcal{I}_l>I_\mathrm{th},
\mathcal{I}_{l-1}>I_\mathrm{th}\right\}$ and $\left\{\prod_{m=1}^M
\left(1+\frac{\eta\,X_m}{N_{\!T}}\!\right)>e^{I_\mathrm{th}},
\prod_{m=1}^M
\left(1+\frac{\eta\,Y_m}{N_{\!T}}\!\right)>e^{I_\mathrm{th}}\right\}$,
have the same probability, which results in
(\ref{eq:varphi_MIMO_IMI}).
\end{proof}

Although (\ref{eq:phi_MIMO_IMI}) and (\ref{eq:varphi_MIMO_IMI}) can
be used to calculate the LCR and AOD of MIMO IMI for small $M$'s,
e.g., $M=2$, via numerical multiple integrals, it is impractical for
large $M$'s. Fortunately, we can approximate $\mathcal{I}_l$ as a
Gaussian random variable for large $M$'s and $N$'s, which is
summarized in the following proposition.

\begin{proposition}\label{prop:MIMO_IMI_Gaussian_Approx} If $M$ and $N$
are large, $\mathcal{I}_l$ can be approximated as a Gaussian random
variable with mean
$\mu_{\!\mathcal{I}}=\mathbb{E}\left[\mathcal{I}_l\right]$ and
variance
$\sigma_{\!\mathcal{I}}^2=\mathbb{E}\left[\mathcal{I}_l^2\right]
-\{\mathbb{E}\left[\mathcal{I}_l\right]\}^2$, where
$\mathbb{E}\left[\mathcal{I}_l\right]$ and
$\mathbb{E}\left[\mathcal{I}_l^2\right]$ are given by
(\ref{eq:1stMoment_MIMO_IMI}) and (\ref{eq:2ndMoment_MIMO_IMI}),
respectively\cite{IEEE_sw27:Hochwald04_Rate_Scheduling,
IEEE_sw27:WangZD04_MIMO_IMI, IEEE_sw27:Smith04_MIMO_IMI_Approx}.
Moreover, we approximate $\mathcal{I}_l$ and $\mathcal{I}_{l-i}$ by
a bivariate Gaussian random vector with mean
$\left(\mathbb{E}\left[\mathcal{I}_l\right],
\mathbb{E}\left[\mathcal{I}_l\right]\right)^T$ and the covariance
matrix
$\Sigma_{\!\mathcal{I}}=\sigma_{\!\mathcal{I}}^2\left(\begin{smallmatrix}1&
\rho_{\!\mathcal{I}}(i)\\\rho_{\!\mathcal{I}}(i)&
1\end{smallmatrix}\right)$, where $\rho_{\!\mathcal{I}}(i)$ is
presented in (\ref{eq:Def_Coeff_MIMO_IMI}).
\end{proposition}

Based on Proposition \ref{prop:MIMO_IMI_Gaussian_Approx}, we have
the following theorem for the LCR and AOD of MIMO IMI.
\begin{theorem}\label{theo:LCR_AOD_MIMO_IMI_Gaussian_Approx}
Using the Gaussian approximation, we can express the LCR and AOD of
MIMO IMI as
\begin{align}
N_{\!\mathcal{I}}(I_\mathrm{th})&=\frac{1}{\pi T_{\!s}}\int_
{\frac{\pi}{4}+\frac{\arcsin\left[\rho_{\!\mathcal{I}}(1)\right]}
{2}}^{\frac{\pi}{2}} \exp\!\!\left(\!-\frac{\tilde{I}_\mathrm{th}^2}
{2\sin^2\theta}\!\right)\!d\theta,
\label{eq:LCR_MIMO_IMI_Gaussian_Approx}\\
\overline{t}_{\!\mathcal{I}}(I_\mathrm{th})&=
\frac{1-Q\!\left(\tilde{I}_\mathrm{th}\right)}
{N_{\!\mathcal{I}}(I_\mathrm{th})},
\label{eq:AOD_MIMO_IMI_Gaussian_Approx}
\end{align} where $\tilde{I}_\mathrm{th}=\frac{I_\mathrm{th}
-\mu_{\!\mathcal{I}}}{\sigma_{\!\mathcal{I}}}$ is the normalized
threshold, and $Q(x)=\frac{1}{\sqrt{2\pi}}\int_x^\infty
e^{-\frac{t^2}{2}}dt$ is the Gaussian $Q$-function.
\end{theorem}
\begin{proof}
See Appendix \ref{app:LCR_AOD_MIMO_IMI_Gaussian_Approx}.
\end{proof}

Theorem \ref{theo:LCR_AOD_MIMO_IMI_Gaussian_Approx} requires
$\mu_{\!\mathcal{I}}$, $\sigma^2_{\!\mathcal{I}}$ and
$\rho_{\!\mathcal{I}}(1)$, which can be obtained from
(\ref{eq:1stMoment_MIMO_IMI}), (\ref{eq:2ndMoment_MIMO_IMI}) and
(\ref{eq:Def_Coeff_MIMO_IMI}). However, for low and high SNRs, we
may use their corresponding approximations. For high SNRs, they are
given by (\ref{eq:1stMoment_MIMO_IMI_HighSNR}),
(\ref{eq:Variance_MIMO_IMI_HighSNR}) and
(\ref{eq:Coeff_HighSNR_MIMO_IMI}), whereas for low SNRs we have
$\mu_{\!\mathcal{I}}=\eta N_{\!R}$,
$\sigma^2_{\!\mathcal{I}}=\frac{\eta^2
N_{\!R}}{N_{\!T}}$\cite{IEEE_sw27:Wang05_IT}, and
$\rho_{\!\mathcal{I}}(1)=\varrho_1^2$, obtained from
(\ref{eq:Coeff_LowSNR_MIMO_IMI}). In practice, the LCR and AOD at
$\mu_{\!\mathcal{I}}$, the ergodic capacity, are of interest, which
simplify Theorem \ref{theo:LCR_AOD_MIMO_IMI_Gaussian_Approx}
considerably.
\begin{corollary}\label{coro:Mean_LCR_AOD_MIMO_IMI_Gaussian_Approx}
The LCR and AOD of MIMO IMI at the level $\mu_{\!\mathcal{I}}$ are,
respectively, given by
\begin{align}
N_{\!\mathcal{I}}(\mu_{\!\mathcal{I}})=\frac{\pi
-2\arcsin\left[\rho_{\!\mathcal{I}}(1)\right]}{4\pi T_{\!s}},
\label{eq:Mean_LCR_MIMO_IMI_Gaussian_Approx}\\
\overline{t}_{\!\mathcal{I}}(\mu_{\!\mathcal{I}})=\frac{2\pi
T_{\!s}} {\pi-2\arcsin\left[\rho_{\!\mathcal{I}}(1)\right]}.
\label{eq:Mean_AOD_MIMO_IMI_Gaussian_Approx}
\end{align}
\end{corollary}
\section{Numerical Results and Discussion}
\label{sec:Numerical_Results_Discussion} 
\begin{figure}[btp]
\centering
\includegraphics[width=0.75\linewidth]
{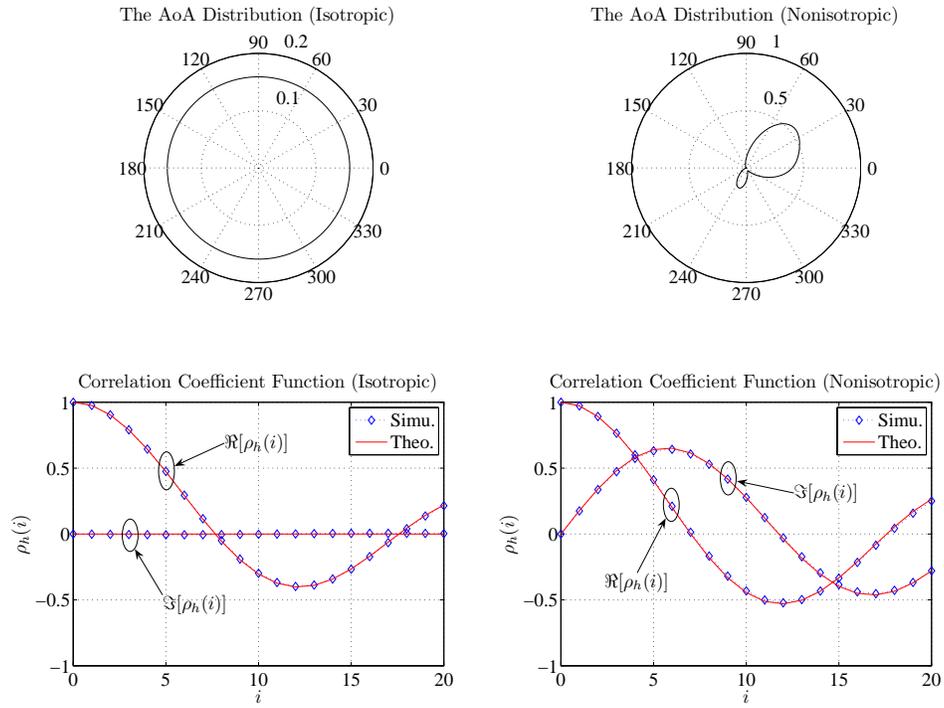}\caption{The AoA distributions
for two scattering examples and the corresponding channel
correlation coefficients.}\label{fig:AoA_Coeff_Plots_Iso_NonIso}
\end{figure}

In this paper, a generic
power spectrum\cite[(8)]{IEEE_sw27:Wang05_LCR_AFD_SISO_Capacity}
\cite{IEEE_sw27:Wang05_IT} is used to simulate time-varying Rayleigh
flat fading channels with \emph{nonisotropic} scattering, according
to the spectral method\cite{IEEE_sw27:Acolatse03}. Similar to
\cite{IEEE_sw27:Wang05_IT}, to verify the accuracy of the derived
formulas, we consider two types of scattering environments:
\emph{isotropic} scattering and \emph{nonisotropic} scattering with
three clusters of scatterers. For \emph{nonisotropic} scattering,
parameters of the three clusters are given by $[P_1, \kappa_1,
\theta_1]=\left[\frac{1}{3}, 6, 0\right]$, $[P_2, \kappa_2,
\theta_2]=\left[\frac{1}{2}, 6, \frac{\pi}{4}\right]$, and $[P_3,
\kappa_3, \theta_3]=\left[\frac{1}{6}, 8, \frac{25\pi}{18}\right]$,
respectively. In addition, in all the simulations, the maximum
Doppler frequency $f_{\!D}$ is set to $10$Hz, and
$T_{\!s}=\frac{1}{20f_{\!D}}$ seconds. The AoA distributions and the
corresponding channel correlation coefficients for the above two
scattering environments are plotted in Fig.
\ref{fig:AoA_Coeff_Plots_Iso_NonIso}.

In the following subsections, Monte Carlo simulations are performed
to verify NACF, the correlation coefficient, LCR and AFD of
\emph{eigen}-channels and the MIMO IMI of two MIMO systems in the
above two propagation environments: one is $4\times4$ and the other
is $12\times3$. The NACF and the correlation coefficient bear almost
the same information. The same comment applies to LCR and AFD.
Therefore, we only report the simulation results for the correlation
coefficient and the LCR, to save space.

\subsection{Eigen-Channels} In this subsection, the correlation
coefficient and the LCR of eigen-channels are considered for both
isotropic and nonisotropic scattering environments.
\subsubsection{Isotropic Scattering} This is Clarke's
model\cite{IEEE_sw27:JakesBook94}, with uniform AoA. The comparison
between the simulation and theoretical results is given in Fig.
\ref{fig:Coeff_LCR_Eigen_4x4_12x3_Iso}.
\subsubsection{Nonisotropic Scattering}
This is a general case, with an arbitrary AoA
distribution\cite{IEEE_sw27:Wang05_LCR_AFD_SISO_Capacity,
IEEE_sw27:Wang05_IT}. The comparison results are presented in Fig.
\ref{fig:Coeff_LCR_Eigen_4x4_12x3_NonIso}.

\begin{figure}[tbp]
\centering
\includegraphics[width=0.75\linewidth]
{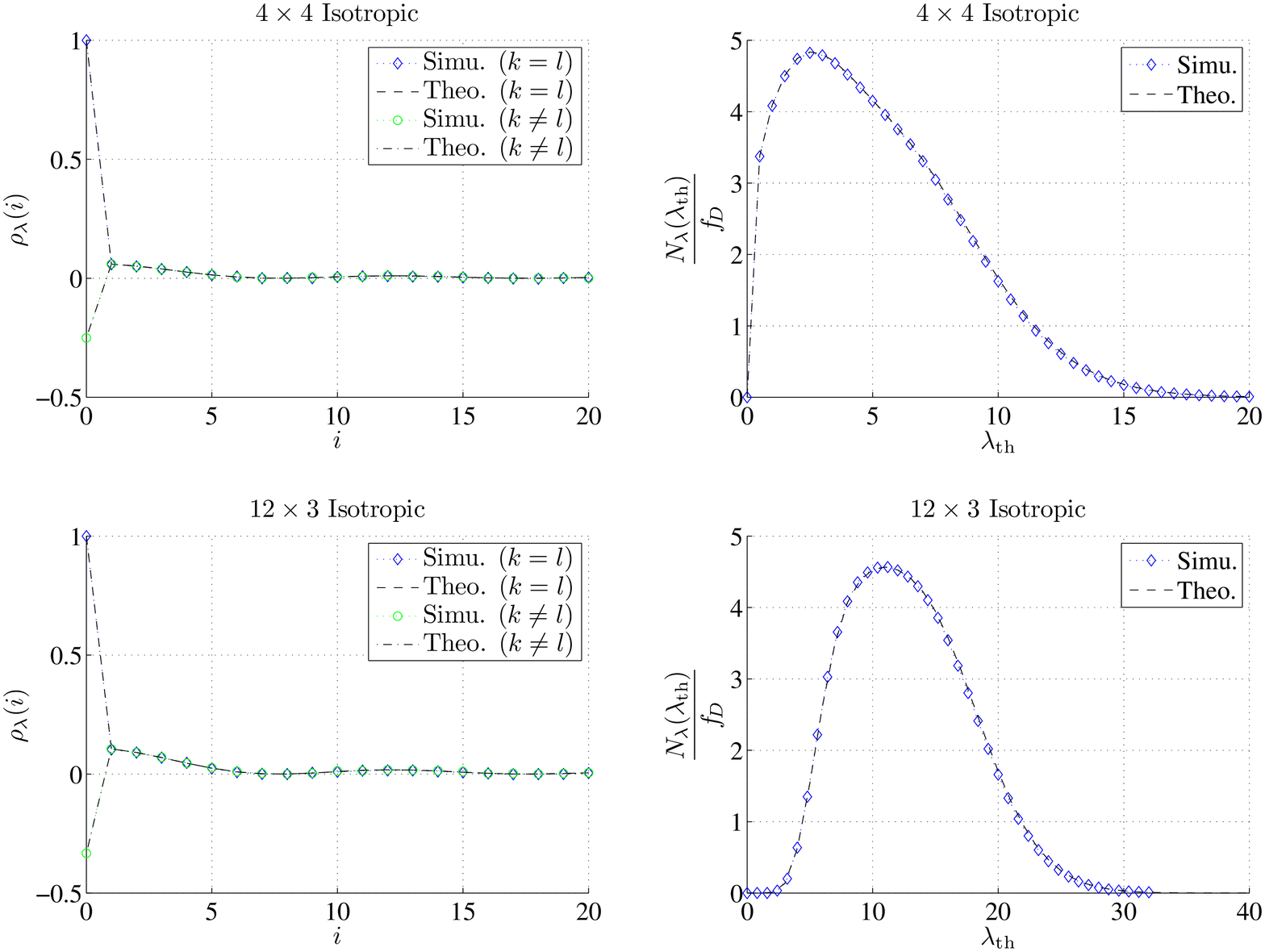} \caption{The correlation
coefficient and the LCR of an \emph{eigen}-channel, in $4\times4$
and $12\times3$ MIMO systems with \emph{isotropic}
scattering.}\label{fig:Coeff_LCR_Eigen_4x4_12x3_Iso}
\end{figure}

\begin{figure}[tp]
\centering
\includegraphics[width=0.75\linewidth]
{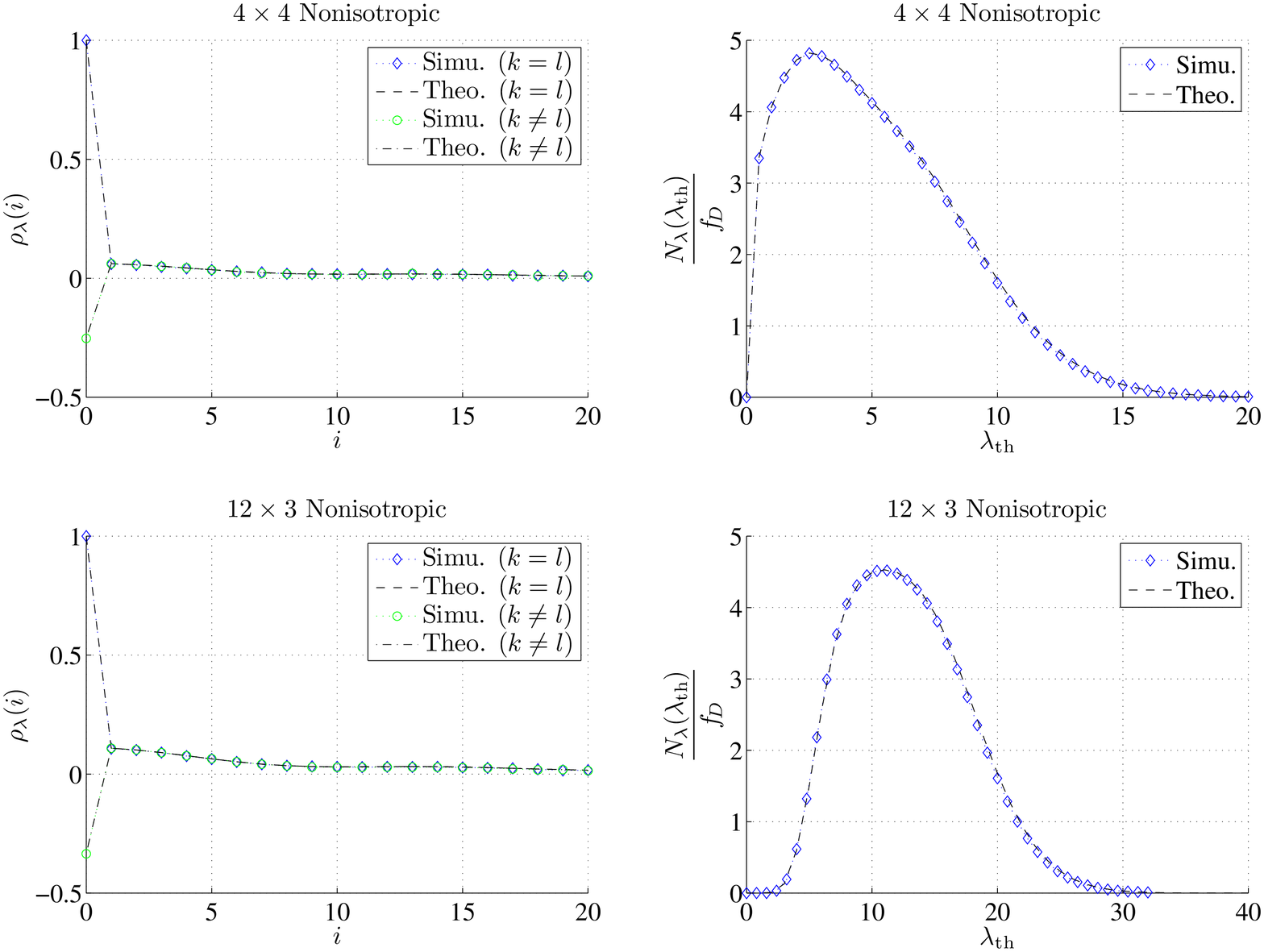} \caption{The correlation
coefficient and the LCR of an \emph{eigen}-channel, in $4\times4$
and $12\times3$ MIMO systems with \emph{nonisotropic}
scattering.}\label{fig:Coeff_LCR_Eigen_4x4_12x3_NonIso}
\end{figure}

In Figs. \ref{fig:Coeff_LCR_Eigen_4x4_12x3_Iso} and
\ref{fig:Coeff_LCR_Eigen_4x4_12x3_NonIso}, the upper left and right
subfigures show the correlation coefficient and the LCR of
eigen-channels in a $4\times4$ MIMO system, respectively, whereas
the lower left and right subfigures show the results in a
$12\times3$ MIMO system. In all figures, ``Simu.'' means simulation.
In the correlation coefficient plots, ``Theo.'' means they are
calculated according to (\ref{eq:Coeff_Eigen-Channel}), and
``$(k=l)$'' denotes the autocorrelation coefficient, whereas
``$(k\neq l)$'' indicates the cross-correlation coefficient. In the
LCR plots, ``Theo'' indicates that the curve is computed using
(\ref{eq:Def_LCR_Up_DT})-(\ref{eq:varrho_1}).

Based on the plots in Figs. \ref{fig:Coeff_LCR_Eigen_4x4_12x3_Iso}
and \ref{fig:Coeff_LCR_Eigen_4x4_12x3_NonIso}, we can see that the
derived analytical formulas perfectly match Monte Carlo simulations.
\subsection{MIMO IMI}
In this subsection, the correlation coefficient and the LCR of MIMO
IMI are presented for both isotropic and nonisotropic scattering
environments at low- and high-SNR regimes. In the simulations and
theoretical calculations, we set $\eta=-20$ dB for low SNR, and
$\eta=30$ dB for high SNR.
\subsubsection{Isotropic Scattering} For this case, the comparison
results are shown in Fig. \ref{fig:Coeff_LCR_MIMO_IMI_4x4_12x3_Iso}.
\subsubsection{Nonisotropic Scattering} The comparison
results regarding nonisotropic scattering are given in Fig.
\ref{fig:Coeff_LCR_MIMO_IMI_4x4_12x3_NonIso}.

\begin{figure}[bp]
\centering
\includegraphics[width=0.75\linewidth]
{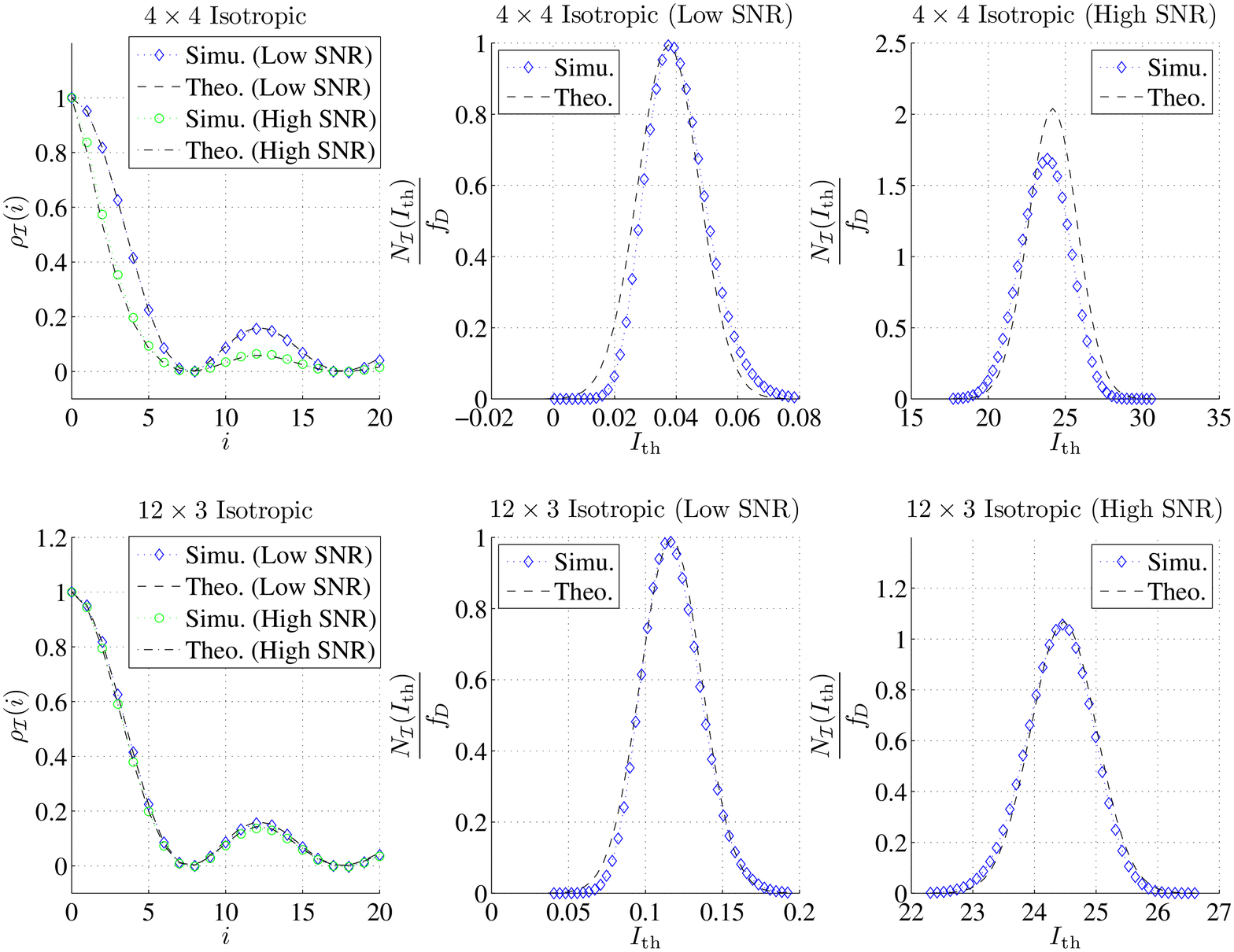} \caption{The
correlation coefficient and the LCR of the MIMO IMI at low- and
high-SNR regimes, in $4\times4$ and $12\times3$ MIMO systems with
\emph{isotropic}
scattering.}\label{fig:Coeff_LCR_MIMO_IMI_4x4_12x3_Iso}
\end{figure}

\begin{figure}[tbp]
\centering
\includegraphics[width=0.75\linewidth]
{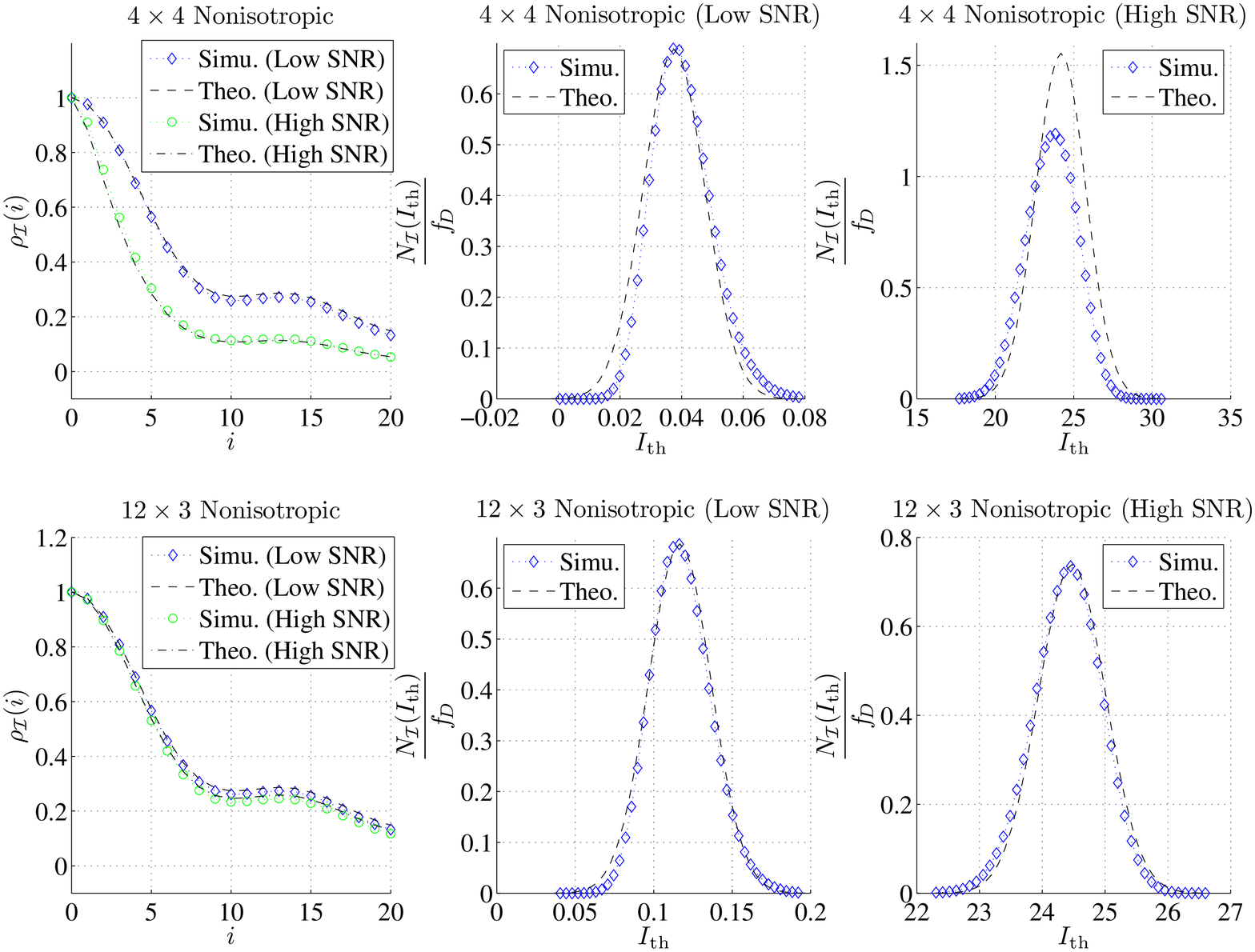} \caption{The
correlation coefficient and the LCR of the MIMO IMI at low- and
high-SNR regimes, in $4\times4$ and $12\times3$ MIMO systems with
\emph{nonisotropic}
scattering.}\label{fig:Coeff_LCR_MIMO_IMI_4x4_12x3_NonIso}
\end{figure}

In Figs. \ref{fig:Coeff_LCR_MIMO_IMI_4x4_12x3_Iso} and
\ref{fig:Coeff_LCR_MIMO_IMI_4x4_12x3_NonIso}, the upper three
subfigures present the correlation coefficient and the LCR of the
MIMO IMI in a $4\times4$ system. Specifically, the upper left
subfigure shows the correlation coefficient at low- and high-SNR
regimes, the upper middle subfigure gives the LCR of the MIMO IMI at
the low-SNR regime, whereas the upper right gives the LCR at the
high-SNR regime. In addition, the lower three subfigures present the
corresponding results in the $12\times3$ system. In the correlation
coefficient plots, ``Theo. (Low SNR)'' corresponds to
(\ref{eq:Coeff_LowSNR_MIMO_IMI}), whereas ``Theo. (High SNR)''
corresponds to (\ref{eq:Coeff_HighSNR_MIMO_IMI}). In the LCR plots,
``Theo.'' means the values are computed from
(\ref{eq:LCR_MIMO_IMI_Gaussian_Approx}), where we used low- and
high-SNR approximations for the mean $\mu_{\!\mathcal{I}}$ and
variance $\sigma^2_{\!\mathcal{I}}$, listed immediately after
Theorem \ref{theo:LCR_AOD_MIMO_IMI_Gaussian_Approx}.

\begin{figure}[tp]
\centering
\includegraphics[width=0.75\linewidth]
{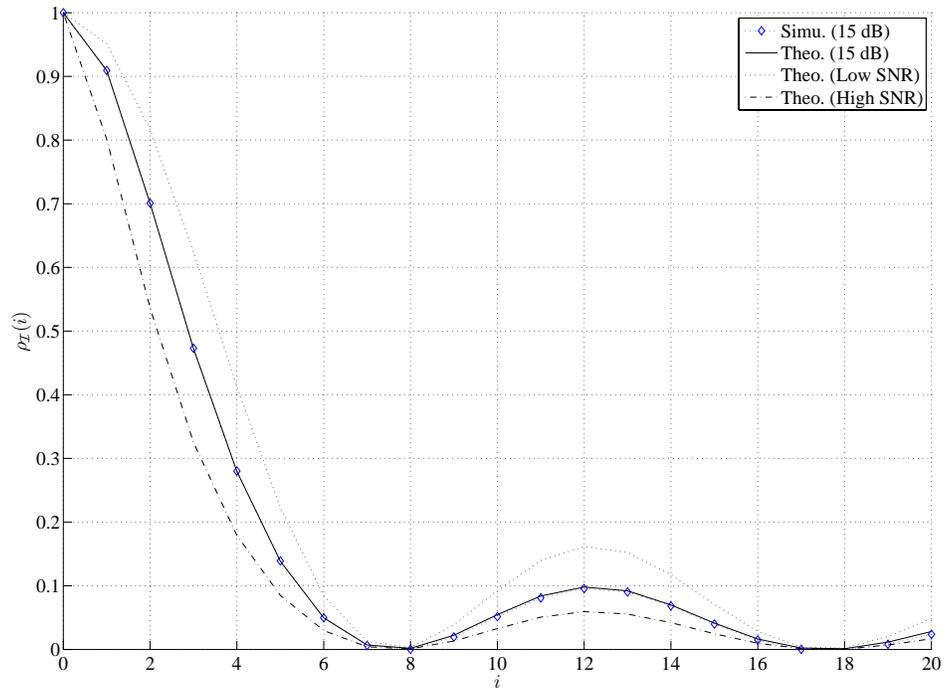} \caption{The correlation coefficient of
the MIMO IMI at $\eta=15$ dB (moderate SNR), in a $4\times4$ system
with \emph{isotropic} scattering.}\label{fig:4x4_Coeff_IMI_15dB}
\end{figure}

From Figs. \ref{fig:Coeff_LCR_MIMO_IMI_4x4_12x3_Iso} and
\ref{fig:Coeff_LCR_MIMO_IMI_4x4_12x3_NonIso}, the following
observations can be made.
\begin{itemize}
\item Correlation coefficient:
If $\nu=\max(N_{\!T}, N_{\!R})-\min(N_{\!T}, N_{\!R})$ is large
compared to $M=\min(N_{\!T}, N_{\!R})$, we can approximate the
correlation coefficient of MIMO IMI by the squared amplitude of the
channel correlation coefficient for all SNRs, since low- and
hign-SNR approximations are very close to each other (see the
results for the $12\times3$ system). However, if $\nu$ is small
compared to $M$, the gap between the low- and high-SNR
approximations is large (see the results for the $4\times4$ system).
Therefore, we need to resort to the exact formulas in
(\ref{eq:2ndMoment_MIMO_IMI}), (\ref{eq:1stMoment_MIMO_IMI}),
(\ref{eq:ACF_MIMO_IMI}) and (\ref{eq:Def_Coeff_MIMO_IMI}) to
calculate the accurate values of the correlation coefficient, for
not so small or large SNRs. For example, at $\eta=15$ dB, the
simulation and exact theoretical curves, as well as low- and
high-SNR approximations are shown in Fig.
\ref{fig:4x4_Coeff_IMI_15dB}, for the correlation coefficient of the
MIMO IMI in a $4\times4$ system.
\item LCR: The Gaussian approximation works well at both low and
high SNRs in large MIMO systems, e.g., the considered $12\times3$
channel. But it is not the case in small MIMO systems, say
$4\times4$, where the Gaussian approximation has an obvious
deviation from the simulation result at high SNR. This is because
the central limit theorem does not hold for IMI in small MIMO
systems\footnote{In fact, there are obvious differences between the
true PDF and the Gaussian approximation in Fig. 1 of
\cite{IEEE_sw27:WangZD04_MIMO_IMI} at $\eta=15$ dB. Larger
deviations are also observed at higher SNRs, say, $\eta=20$ dB, in
Fig. 1 of \cite{IEEE_sw27:WangZD04_MIMO_IMI}.}. For this case, we
can numerically compute the multiple integrals given in
(\ref{eq:phi_MIMO_IMI}) and (\ref{eq:varphi_MIMO_IMI}), to calculate
the LCR.
\item LCR: Compared Fig. \ref{fig:Coeff_LCR_Eigen_4x4_12x3_Iso} and
Fig. \ref{fig:Coeff_LCR_Eigen_4x4_12x3_NonIso}, we find the LCR of
an eigen-channel is not sensitive to the scattering environment,
which is not the case for the LCR of MIMO IMI. Furthermore, based on
Figs. \ref{fig:Coeff_LCR_MIMO_IMI_4x4_12x3_Iso} and
\ref{fig:Coeff_LCR_MIMO_IMI_4x4_12x3_NonIso}, we can see that the
IMI in a nonisotropic scattering environment has less fluctuations
than that in the isotropic scattering scenario.
\end{itemize}
\section{Conclusion}
\label{sec:conclusion} In this paper, closed-form expressions for
several key second-order statistics such as the autocorrelation
function, the correlation coefficient, level crossing rate and
average fade/outage duration of \emph{eigen}-channels and the
instantaneous mutual information (IMI) are derived in MIMO
time-varying Rayleigh flat fading channels.

Simulation and analytical results show that the eigen-modes tend to
be spatio-temporally uncorrelated in large MIMO systems, and the
correlation coefficient of the IMI can be well approximated by the
squared amplitude of the correlation coefficient of the channel, if
the difference between the number of Tx and Rx antennas is much
larger than the minimum number of Tx and Rx antennas. In addition,
we have also observed that the LCR of an eigen-mode is less
sensitive to the scattering environment than the IMI.

The analytical expressions, supported by Monte Carlo simulations,
provide quantitative information regarding the dynamic behavior of
MIMO channels. They also serve as useful tools for MIMO system
designs. For example, one may improve the performance of the
feedbacked-IMI-based rate scheduler in a multiuser MIMO system by
exploiting the temporal correlation of the IMI of each user.

\appendices
\section{Proof of Lemma \ref{lem:Moments_Eigen-Channel}}
\label{app:Proof_Moments_Eigen-Channel} Although the mean and second
moment of $\lambda_m(l)$ were respectively given by (57) and (58) in
\cite{IEEE_sw27:Hochwald04_Rate_Scheduling} via a smart indirect
method, we calculate them directly using its marginal PDF in
(\ref{eq:PDF_x}), as follows.

Using 8.902.2\cite[pp. 1043]{IEEE_sw27:RyzhikBook_5th}, we can
rewrite (\ref{eq:PDF_x}) as
\begin{equation}\label{eq:PDF_x_alternate_way}
p(x)\!=\!\!\frac{(M\!-\!1)!x^\nu}{(M\!+\!\nu\!-\!1)!
e^x}\!\left\{\!\!\left[L_{\!M\!-\!1}^{\nu}\!(x)\right]^\prime
\!L_{\!M}^\nu\!(x)\!-\!\left[L_{\!M}^{\nu}\!(x)\right]^\prime
\!L_{\!M\!-\!1}^\nu\!(x)\!\right\},
\end{equation} where $^\prime$ mean the derivative with respect to
$x$. With\cite[pp. 1062, 8.971.2]{IEEE_sw27:RyzhikBook_5th}
\begin{equation}
\left[L_n^\nu(x)\right]^\prime=-L_{n-1}^{\nu+1}(x)
\end{equation} and\cite[pp. 1062, 8.971.5]{IEEE_sw27:RyzhikBook_5th}
\begin{equation}\label{eq:Laguerre_Ploy_Relation}
L_n^k(x)=L_n^{k+1}(x)-L_{n-1}^{k+1}(x),
\end{equation}
(\ref{eq:PDF_x_alternate_way}) further reduces to
\begin{equation}\label{eq:PDF_x_alternate_way_1}
p(x)\!=\!\!\frac{(M\!-\!1)!x^\nu}{(M\!+\!\nu\!-\!1)! e^x}
\!\left\{\!\!\left[L_{\!M\!-\!1}^{\nu+1}(x)\right]^{\!2}\!-\!
L_{\!M}^{\nu+1}(x)L_{\!M\!-\!2}^{\nu+1}(x)\!\right\}\!\!,
\end{equation} where the convention $L_m^k(x)=0, m<0$ should be used
when it is applicable.

Using (\ref{eq:PDF_x_alternate_way_1}), we obtain
$\mathbb{E}[\lambda_m(l)]$ as
\begin{equation}\label{eq:1stMoment_Eigen-Channel_Proof}
\begin{split}
\mathbb{E}[\lambda_m(l)]&=\int_0^\infty xp(x)dx,\\
&=\frac{(M\!-\!1)!}{(M\!+\!\nu\!-\!1)!}\bigg\{\int_0^\infty
x^{\nu+1} e^{-x}\left[L_{\!M\!-\!1}^{\nu+1}(x)\right]^{\!2}dx-
\int_0^\infty x^{\nu+1}
e^{-x}L_{\!M}^{\nu+1}(x)L_{\!M\!-\!2}^{\nu+1}(x)dx \bigg\},\\
&=M+\nu,
\end{split}
\end{equation}
where the orthogonality of Laguerre polynomials \cite[pp.~267,
7.414.3]{IEEE_sw27:BeckmannBook73} is used, i.e.
\begin{equation}\label{eq:Orthorgonality_Laguerre_Poly}
\int_0^\infty e^{-x}x^\nu L_k^\nu(x)L_l^\nu(x)
=\frac{(k+\nu)!}{k!}\delta_{k,l}.
\end{equation} The last line results in (\ref{eq:1stMoment_Eigen-Channel}),
considering $N=M+\nu$.

By substituting (\ref{eq:Laguerre_Ploy_Relation}) with $k=\nu+1$
into (\ref{eq:PDF_x_alternate_way_1}) and using
(\ref{eq:Orthorgonality_Laguerre_Poly}), we can easily obtain
(\ref{eq:2ndMoment_Eigen-Channel}).

\section{Proof of Lemma \ref{lem:ACF_Eigen-Channel}}
\label{app:Proof_ACF_Eigen-Channel}\subsection{The case of $i=0$}
For $i=0$, the value of $\mathbb{E}[\lambda^2_m(l)]$ is given in
(\ref{eq:2ndMoment_Eigen-Channel}).
\subsection{The case of $i\neq0$}
For $i\neq0$, we need the following two lemmas.
\begin{lemma}\label{lem:Integral_Laguerres}
While $j$, $k$ and $\nu$ are non-negative integers, the value of the
integral,\\$I_1(j,k,\nu)=\int_0^\infty
\!x^{\nu+1}e^{-x}L_j^\nu(x)L_k^\nu(x)dx$, is given by
\begin{equation}\label{eq:Integral_Laguerres}
I_1(j,k,\nu)\!=\!
\begin{cases}\frac{(2k+\nu+1)(k+\nu)!}{k!},&|j-k|=0,\\
-\frac{\left[\min(j,k)+\nu+1\right]!}{\left[\min(j,k)\right]!}, &|j-k|=1,\\
0, &|j-k|\geq2.
\end{cases}
\end{equation}
\end{lemma}
\begin{proof}
Using (\ref{eq:Laguerre_Ploy_Relation}), we have
\begin{equation}\label{eq:LjLk}
L_j^\nu(x)L_k^\nu(x)=L_j^{\nu+1}(x)L_k^{\nu+1}(x)
+L_{j-1}^{\nu+1}(x)L_{k-1}^{\nu+1}(x)
-L_j^{\nu+1}(x)L_{k-1}^{\nu+1}(x)-L_{j-1}^{\nu+1}(x)L_k^{\nu+1}(x).
\end{equation} Substitution of (\ref{eq:LjLk}) into $I_1(j,k,\nu)$
results in (\ref{eq:Integral_Laguerres}), with the aid of
(\ref{eq:Orthorgonality_Laguerre_Poly}) and the convention
$L_m^k(x)=0, m<0$.
\end{proof}

\begin{lemma}\label{lem:jpdf_beta_alpha_alternate_way}
The joint PDF in (\ref{eq:jpdf_beta_alpha}) can be written in the
following equivalent form
\begin{equation}\label{eq:jpdf_beta_alpha_alternate_way}
p(x,y)=\frac{1}{M^2}\sum_{j=M}^\infty\sum_{k=0}^{M-1}
\bigg[\frac{j!k!}{(j+\nu)!(k+\nu)!} \varrho_i^{2(j-k)}\frac{x^\nu
y^\nu}{e^{x+y}}
L_j^\nu(x)L_j^\nu(y)L_k^\nu(x)L_k^\nu(y)\bigg]+p(x)p(y),
\end{equation} where $p(\cdot)$ is the marginal PDF given by
(\ref{eq:PDF_x}).
\end{lemma}
\begin{proof}
By applying the Hille-Hardy formula\cite[pp.~185,
(46)]{IEEE_sw27:BeckmannBook73}
\begin{equation}\label{eq:Hille-Hardy-Formula}
\sum_{k=0}^\infty\frac{k!z^k}{(k+\nu)!}L_k^\nu(x)L_k^\nu(y)
=\frac{(xyz)^{-\frac{\nu}{2}}}{1-z}\exp\left(-z\frac{x+y}{1-z}\right)
I_\nu\!\!\left(\frac{2\sqrt{xyz}}{1-z}\right), |z|<1,
\end{equation} to (\ref{eq:jpdf_beta_alpha}), we can obtain
(\ref{eq:jpdf_beta_alpha_alternate_way}) after some algebraic
manipulations.
\end{proof}

Using Lemmas \ref{lem:Integral_Laguerres} and
\ref{lem:jpdf_beta_alpha_alternate_way}, it is straightforward to
obtain
\begin{equation}
r_{m,m}(i)=\frac{1}{M^2}\frac{M!(M\!-\!1)!\varrho_i^2}
{(M\!+\!\nu)!(M\!+\!\nu\!-\!1)!}
\left[I_1(M,M\!-\!1,\nu)\right]^2\!+\!N^2,
\end{equation} which reduces to (\ref{eq:ACF_Eigen-Channel}), based
on (\ref{eq:Integral_Laguerres}) and $N=M+\nu$.
\section{Proof of Lemma \ref{lem:CCF_Eigen-Channel}}
\label{app:Proof_CCF_Eigen-Channel}
\subsection{The case of $i=0$} For $i=0$, we
need the following proposition.
\begin{proposition}\label{prop:jpdf_x1_x2}
If $(x_1,x_2)$ are a pair of eigenvalues, randomly selected from
$\{\lambda_m(l)\}_{m=1}^M$, then their joint PDF is given
by\cite{IEEE_sw27:Wang_SIAM}
\begin{equation}\label{eq:jpdf_x1_x2}
p(x_1,x_2)\!=\!\frac{(x_1x_2)^\nu
e^{-(x_1+x_2)}}{M(M-1)}\!\sum_{\begin{subarray}{c}p,q=0\\p\neq
q\end{subarray}}^{M-1}\!\frac{p!q!}{(p+\nu)!(q+\nu)!}
\left\{\!\left[L_p^\nu(x_1)L_q^\nu(x_2)\right]^2
\!-\!L_p^\nu(x_1)L_q^\nu(x_1)L_p^\nu(x_2)L_q^\nu(x_2)\!\right\}\!.
\end{equation}
\end{proposition}

Note that (\ref{eq:jpdf_x1_x2}) is different from
(\ref{eq:jpdf_beta_alpha}). By reordering the items, we can rewrite
(\ref{eq:jpdf_x1_x2}) as
\begin{equation}\label{eq:jpdf_x1_x2_alternate_way}
p(x_1,x_2)\!=\frac{M}{M-1}p(x_1)p(x_2)- \frac{(x_1x_2)^\nu
e^{-(x_1+x_2)}}{M(M-1)}\sum_{j=0}^{M-1}\!
\sum_{k=0}^{M-1}\!\frac{j!k!}{(j+\nu)!(k+\nu)!}L_j^\nu(x_1)
L_k^\nu(x_1)L_j^\nu(x_2)L_k^\nu(x_2).
\end{equation}

Using Lemma \ref{lem:Integral_Laguerres} and
(\ref{eq:jpdf_x1_x2_alternate_way}), it is easy to obtain
\begin{equation}\label{eq:Proof_CCF_Eigen-Channel}
r_{m,n}(i)\!=\!\frac{M}{M\!-\!1}N^2\!-\!\frac{S}{M(M\!-\!1)},
\end{equation} where $S=\sum_{j=0}^{M-1}
\!\sum_{k=0}^{M-1}\!\frac{j!k!\left[I_1(j,k,\nu)\right]^2}{(j+\nu)!(k+\nu)!}$.
According to (\ref{eq:Integral_Laguerres}), we have
\begin{equation}\label{eq:Summation_I_Squared}
\begin{split}
S&\!=\!\sum_{k=0}^{M-1}(2k\!+\!\nu\!+\!1)^2\!
+\!2\sum_{k=0}^{M-2}(k\!+\!1)(k\!+\!\nu\!+\!1),\\
&\!=\!M\!N(M+N-1),
\end{split}
\end{equation} where the last line is derived based on
$\sum_{k=0}^nk=\frac{n(n+1)}{2}$\cite[pp. 2,
0.121.1]{IEEE_sw27:RyzhikBook_5th} and
$\sum_{k=0}^nk^2=\frac{n(n+1)(2n+1)}{6}$\cite[pp. 2,
0.121.2]{IEEE_sw27:RyzhikBook_5th}.

Substitution of (\ref{eq:Summation_I_Squared}) into
(\ref{eq:Proof_CCF_Eigen-Channel}) proves the first part of Lemma
\ref{lem:CCF_Eigen-Channel}, i.e., $i=0$. Note that the same result
was derived in Lemma $A$ of
\cite{IEEE_sw27:Hochwald04_Rate_Scheduling} via an indirect method.
\subsection{The case of $i\neq0$}
Note that $\{\lambda_m(l)\}_{m=1}^M$ and
$\{\lambda_m(l-i)\}_{m=1}^M$ are unordered eigenvalues of
$\mathbf{H}(l)\mathbf{H}^\dag(l)$ and
$\mathbf{H}(l-i)\mathbf{H}^\dag(l-i)$, respectively, for $i\neq0$.
So the bivariate PDF of $\{\lambda_m(l), \lambda_n(l-i)\}$, $m\neq
n$, is the same as that of $\{\lambda_m(l), \lambda_m(l-i)\}$, the
latter given in (\ref{eq:jpdf_beta_alpha}). Therefore,
$r_{m,n}(i)=r_{m,m}(i), i\neq0, \forall m,n$, where
$r_{m,m}(i)=N^2+\frac{N\varrho_i^2}{M}, i\neq0$, is proved in
Appendix \ref{app:Proof_ACF_Eigen-Channel}.
\section{Proof of Lemma \ref{lem:Moments_MIMO_IMI}}
\label{app:Proof_Moments_MIMO_IMI} According to
(\ref{eq:MIMO_IMI_Eigenvalue}), we have
\begin{equation}\label{eq:proof_1stMoment_MIMO_IMI}
\begin{split}
\mathbb{E}[\mathcal{I}_l]&=M\mathbb{E}\left[\ln\left(1+\frac{\eta}
{N_{\!T}}\lambda_m(l)\right)\right],\\
&=M\int_0^\infty \ln\left(1+\frac{\eta} {N_{\!T}}x\right)p(x)dx,
\end{split}
\end{equation} where $p(x)$ is given in (\ref{eq:PDF_x}).
Substitution of (\ref{eq:PDF_x}) and
(\ref{eq:Laguerre_Poly_SeriesFormat}) into
(\ref{eq:proof_1stMoment_MIMO_IMI}) results in
(\ref{eq:1stMoment_MIMO_IMI}), with the aid of the following
integral identity\cite[(67)]{IEEE_sw27:Wang05_IT}
\begin{equation}\label{eq:MeijerG_for_Integrand}
\int_0^\infty \!\!x^ke^{-x}\ln(1\!+\!\omega x)\mathrm{d}x=G_{2,
3}^{3, 1}\!\!\left(\!\frac{1}{\omega}\left|\!\!\!
\begin{array}{c}
0,1 \\
0,0,k\!+\!1
\end{array}\right.
\!\!\!\!\right),
\end{equation} where $G$ is Meijer's $G$ function\cite[pp.~1096,
9.301]{IEEE_sw27:RyzhikBook_5th}.

Similarly, we have
\begin{equation}\label{eq:proof_2ndMoment_MIMO_IMI}
\begin{split}
\mathbb{E}[\mathcal{I}^2_l]&=M\mathbb{E}\left[\ln^2\left(1+\frac{\eta}
{N_{\!T}}\lambda_m(l)\right)\right]
+M(M\!-\!1)\underbrace{\mathbb{E}\!\left[\!\ln\!
\left(\!1\!+\!\frac{\eta}
{N_{\!T}}\lambda_m(l)\!\right)\!\ln\!\left(\!1\!+\!\frac{\eta}
{N_{\!T}}\lambda_n(l)\!\right)\!\right]}_{m\neq n},\\
&=M\int_0^\infty \ln^2\left(1+\frac{\eta}
{N_{\!T}}x\right)p(x)dx+M(M\!-\!1)
\int_0^\infty\!\!\!\!\int_0^\infty \!\ln\!\left(\!1\!+\!\frac{\eta}
{N_{\!T}}x_1\!\right)\!\ln\!\left(\!1\!+\!\frac{\eta}
{N_{\!T}}x_2\!\right)\!p(x_1,x_2)dx_1dx_2,
\end{split}
\end{equation} where $p(x)$ and $p(x_1,x_2)$ are given in (\ref{eq:PDF_x})
and (\ref{eq:jpdf_x1_x2_alternate_way}), respectively. Substitution
of (\ref{eq:PDF_x}), (\ref{eq:Laguerre_Poly_SeriesFormat}) and
(\ref{eq:jpdf_x1_x2_alternate_way}) into
(\ref{eq:proof_2ndMoment_MIMO_IMI}) leads us to
(\ref{eq:2ndMoment_MIMO_IMI}), upon using
(\ref{eq:MeijerG_for_Integrand}) and the following integral
equality\cite[(69)]{IEEE_sw27:Wang05_IT}
\begin{equation}\label{eq:MeijerG_for_Integrand_squared}
\int_0^\infty \!\!x^ke^{-x}\ln^2(1+\omega
x)\mathrm{d}x=2e^\frac{1}{\omega}\omega^{-(k+1)}\sum_{j=0}^k{k
\choose j}(-1)^{k-j}G_{3,4}^{4,0}\!\left(\!\frac{1}{\omega}\!\left|
\begin{array}{c}
 -\!j,-\!j,-\!j \\
 0,-\!j\!-\!1,-\!j\!-\!1,-\!j\!-\!1
\end{array}
\right.\!\!\!\!\right).
\end{equation}
\section{Proof of Theorem \ref{theo:NACF_Coeff_MIMO_IMI_HighSNR_Approx}}
\label{app:NACF_Coeff_MIMO_IMI_HighSNR_Approx} First we derive the
expressions for the first and second moments of $\mathcal{I}_l$ in
(\ref{eq:MIMO_IMI_HighSNR_Approx}), based on the following lemma.
\begin{lemma}\label{lem:1st2ndMoments_log_det_correlated_Wishart}
Let $\mathbf{X}=\left(x_{m,n}\right)$ be a random matrix with $M$
rows and $N$ columns, $M\leq N$, where each element is a zero mean
unit variance complex Gaussian random variable and all the $N$
columns are i.i.d $M$-variate random vectors with the same $M\times
M$ positive definite covariance matrix $\Sigma$. The mean and
variance of $\ln\det\left(\mathbf{X}\mathbf{X}^\dag\right)$ are
\begin{align}
\mathbb{E}\left[\ln\det\left(\mathbf{X}\mathbf{X}^\dag\right)\right]
&=\sum_{m=0}^{M-1}\!\psi_{\!N-m}
+\ln\det\Sigma,\label{eq:1stMoment_Correlated_Wishart}\\
\var\left[\ln\det\left(\mathbf{X}\mathbf{X}^\dag\right)\right]
&=\sum_{m=0}^{M-1}\zeta(2,N-m).\label{eq:2ndMoment_Correlated_Wishart}
\end{align}
\end{lemma}
\begin{proof}
According to Theorem 1.1 of \cite{IEEE_sw27:Goodman63_Det},
$\frac{\det\left(\mathbf{X}\mathbf{X}^\dag\right)}{2^{-M}\det\Sigma}$
has the same distribution as the product of $M$ independent $\chi^2$
random variables with $2N$, $2(N-1)$, $\cdots$, $2(N-M+1)$ degrees
of freedom, respectively. Therefore, we can express
$\ln\det\left(\mathbf{X}\mathbf{X}^\dag\right)$ as
\begin{equation}\label{eq:MIMO_IMI_HighSNR_Approx_ChiSquare}
\ln\det\left(\mathbf{X}\mathbf{X}^\dag\right)
\operatornamewithlimits{=}^\mathrm{d} \sum_{m=0}^{M-1}\ln
\left(\sqrt[M]{\det\!\Sigma}\,y_m\right),
\end{equation}
where the notation
$\displaystyle{\operatornamewithlimits{=}^\mathrm{d}}$ indicates
``equal to in distribution'', $2y_m$ is a $\chi^2$ random variable
with $2(N-m)$ degrees of freedom, and
$\left\{y_m\right\}_{m=0}^{M-1}$ are independent.

Based on the results in \cite{IEEE_sw27:Wang05_IT}, we have
$\mathbb{E}\left[\ln y_m\right]=H_{\!N-m-1}-\mathtt{C}$ and
$\var\left[\ln y_m\right]=\zeta(2,N-m)$, where $H_{\!k}$ is the
$k^\text{th}$ harmonic number \cite[pp. 29,
(2.13)]{IEEE_sw27:Knuth_2nd}, defined by
${H}_{\!k}=\sum_{j=1}^k\frac{1}{j}$ for $k\geq1$ with ${H}_0=0$, and
$\mathtt{C}=0.577215\cdots$ is the Euler-Mascheroni constant
\cite[pp.~\textrm{xxx}]{IEEE_sw27:RyzhikBook_5th}. This completes
the proof if we note $\psi_{\!k+1}=H_{\!k}-\mathtt{C}$\cite[pp. 952,
8.365.4]{IEEE_sw27:RyzhikBook_5th}.

It is interesting to observe that the correlation matrix $\Sigma$
affects the mean of $\ln\det\left(\mathbf{X}\mathbf{X}^\dag\right)$
in (\ref{eq:1stMoment_Correlated_Wishart}), but has no impact on its
variance in (\ref{eq:2ndMoment_Correlated_Wishart}).
\end{proof}

According to Theorem 1.1.2 of \cite{IEEE_sw27:GuptaBook99} we have
\begin{equation}\label{eq:relation_prod_eigen_det}
\prod_{m=1}^M\lambda_m(l)=\begin{cases}
\det\left(\mathbf{H}_l\mathbf{H}_l^\dag\right),N_{\!R}\leq N_{\!T},\\
\det\left(\mathbf{H}_l^\dag\mathbf{H}_l\right),N_{\!R}> N_{\!T}.
\end{cases}
\end{equation} By applying Lemma \ref{lem:1st2ndMoments_log_det_correlated_Wishart}
and (\ref{eq:relation_prod_eigen_det}) to
(\ref{eq:MIMO_IMI_HighSNR_Approx}) with $\Sigma=\mathbf{I}_M$, it is
straightforward to write the mean and variance of $\mathcal{I}_l$ as
\begin{equation}\label{eq:1stMoment_MIMO_IMI_HighSNR}
\mathbb{E}\left[\mathcal{I}_l\right]\approx\sum_{m=0}^{M-1}\!\psi_{\!N-m}
+M\ln\frac{\eta}{N_{\!T}},
\end{equation}
and
\begin{equation}\label{eq:Variance_MIMO_IMI_HighSNR}
\var\left[\mathcal{I}_l\right]\approx\sum_{m=0}^{M-1}\zeta(2,N-m),
\end{equation} respectively. These two are consistent
with the results in \cite{IEEE_sw27:Hochwald04_Rate_Scheduling},
where an implicit complex extension of Theorem 3.3.4 of
\cite{IEEE_sw27:GuptaBook99} was used. Clearly, the second moment of
$\mathcal{I}_l$ is given by
\begin{equation}\label{eq:2ndMoment_MIMO_IMI_HighSNR}
\mathbb{E}\!\left[\mathcal{I}_l^2\right]\!\approx\!
\sum_{m=0}^{M-1}\zeta(2,N\!-\!m)\!+\!\left(\sum_{m=0}^{M-1}\!\psi_{\!N-m}
\!+\!M\ln\!\frac{\eta}{N_{\!T}}\!\right)^{\!\!2}\!\!.
\end{equation}

For calculating the autocorrelation of $\mathcal{I}_l$, we need the
following lemma.
\begin{lemma}\label{lem:Integral_Laguerres_logarithm}
With $j$, $k$ and $\nu$ as non-negative integers and $j\neq k$, the
value of the integral\\$I_2(j,k,\nu)=\int_0^\infty \!\left(\ln
x\right)x^\nu e^{-x}L_j^\nu(x)L_k^\nu(x)dx$ is given by
\begin{equation}\label{eq:Integral_Laguerres_logarithm}
I_2(j,k,\nu)=\frac{\left[\min(j,k)+\nu\right]!}
{\left[\min(j,k)\right]!\left[\min(j,k)-\max(j,k)\right]}.
\end{equation}
\end{lemma}
\begin{proof}
First we consider $j>k$. Substitution of $L_k^\nu(x)$ with
(\ref{eq:Laguerre_Poly_SeriesFormat}) into $I_2(j,k,\nu)$ gives
\begin{equation}\label{eq:I2jknu}
\begin{split}
I_2(j,k,\nu)&=\sum_{p=0}^k{k\!+\!\nu\choose
k\!-\!p}\frac{(-1)^p}{p!}\!\!\int_0^\infty \!\!\left(\ln
x\right)x^{p+\nu} e^{-x}L_j^\nu(x)dx,\\
&=\sum_{p=0}^k{k\!+\!\nu\choose k\!-\!p}\frac{(-1)^p}{p!}
\frac{(-1)^{p-1}p!(j\!-\!p\!-\!1)!(p\!+\!\nu)!}{j!},\\
&=-\frac{(k\!+\!\nu)!}{j!}\sum_{p=0}^k
\frac{(j\!-\!p\!-\!1)!}{(k\!-\!p)!},
\end{split}
\end{equation} where the second line comes from
2.19.6.2\cite[pp.~469]{IEEE_sw27:Prudnikov_Vol2_1st}. Using
$\sum_{q=0}^m {n+q\choose n}={n+m+1\choose n+1}$\cite[pp.~4,
0.151.1]{IEEE_sw27:RyzhikBook_5th}, we have $\sum_{p=0}^k
\frac{(j\!-\!p\!-\!1)!}{(k\!-\!p)!}
{\displaystyle\operatornamewithlimits{=}^{q=k-p}}\sum_{q=0}^k
{j-k-1-q\choose j-k-1}(j-k-1)!={j\choose
j-k}(j-k-1)!=\frac{j!}{k!(j-k)}$, which reduces (\ref{eq:I2jknu}) to
\begin{equation}\label{eq:I2jknu_Final1}
I_2(j,k,\nu)=\frac{(k\!+\!\nu)!}{k!(k-j)}.
\end{equation}

Similarly, for $j<k$, we obtain
\begin{equation}\label{eq:I2jknu_Final2}
I_2(j,k,\nu)=\frac{(j\!+\!\nu)!}{j!(j-k)}.
\end{equation}

Combination of (\ref{eq:I2jknu_Final1}) and (\ref{eq:I2jknu_Final2})
results in (\ref{eq:Integral_Laguerres_logarithm}).
\end{proof}

Now we proceed to prove (\ref{eq:NACF_HighSNR_MIMO_IMI}) and
(\ref{eq:Coeff_HighSNR_MIMO_IMI}). Based on the high-SNR
approximation of $\mathcal{I}_l$ in
(\ref{eq:MIMO_IMI_HighSNR_Approx}), we have
\begin{equation}\label{eq:ACF_MIMO_IMI_HighSNR}
\begin{split}
r_{\!\mathcal{I}}(i) &\approx\sum_{m=1}^{M}\sum_{n=1}^{M}\mathbb{E}
\left[\ln\frac{\eta \,\lambda_m(l)}{N_{\!T}}\ln\frac{\eta
\,\lambda_n(l-i)}{N_{\!T}}\right],\\
&=M^2\,\mathbb{E}\!\left[\ln\frac{\eta
\,\lambda_1(l)}{N_{\!T}}\ln\frac{\eta
\,\lambda_1(l-i)}{N_{\!T}}\right],\\
&=M^2\left[\ln^2\frac{\eta}{N_{\!T}}+2\ln\frac{\eta}{N_{\!T}}
\mathbb{E}[\ln\lambda(l)]+r_{\ln\!\lambda}(i)\right],
\end{split}
\end{equation}
where $r_{\ln\!\lambda}(i)=\mathbb{E}\left[\ln\lambda(l)
\ln\lambda(l-i)\right]$. Using
(\ref{eq:jpdf_beta_alpha_alternate_way}) and Lemma
\ref{lem:Integral_Laguerres_logarithm}, $r_{\ln\!\lambda}(i)$ can be
evaluated as
\begin{equation}\label{eq:ACF_ln_lambda}
r_{\ln\!\lambda}(i)=\frac{1}{M^2}\sum_{j=M}^\infty\sum_{k=0}^{M-1}
\frac{j!k!\varrho_i^{2(j-k)}I_2^2(j,k,\nu)}{(j+\nu)!(k+\nu)!}
+\left\{\mathbb{E}[\ln\lambda(l)]\right\}^2.
\end{equation}

By substituting (\ref{eq:Integral_Laguerres_logarithm}) and
(\ref{eq:ACF_ln_lambda}) into (\ref{eq:ACF_MIMO_IMI_HighSNR}), we
obtain
\begin{equation}\label{eq:ACF_MIMO_IMI_HighSNR_1}
r_{\!\mathcal{I}}(i)\approx\sum_{k=0}^{M-1}S(k,\nu,\varrho_i)
+\left\{\mathbb{E}[\mathcal{I}_l]\right\}^2,
\end{equation} where $S(k,\nu,\varrho_i)=\sum_{j=M}^\infty
\frac{j!(k+\nu)!\varrho_i^{2(j-k)}}{k!(j+\nu)!(j-k)^2}$, and
$\mathbb{E}[\mathcal{I}_l]\approx M\mathbb{E}\left[\ln\frac{\eta
\,\lambda(l)}{N_{\!T}}\right]$ is approximated by
(\ref{eq:1stMoment_MIMO_IMI_HighSNR}). By introducing a new variable
$p=j-M$ in $S(k,\nu,\varrho_i)$ and using the Pochhammer symbol
$(x)_n=x(x+1)\cdots(x+n-1)$, we can rewrite $S(k,\nu,\varrho_i)$ as
\begin{equation}\label{eq:S(k,nv,varrho)}
\begin{split}
S(k,\nu,\varrho_i)&=\frac{M!(k+\nu)!\varrho_i^{2(M-k)}}{(M-k)^2N!k!}
\sum_{p=0}^\infty
\frac{\left[(M-k)_p\right]^2(M+1)_p(1)_p}{\left[(M-k+1)_p\right]^2(N+1)_p}
\frac{\varrho_i^{2p}}{p!},\\
&=\frac{M!(k+\nu)!\varrho_i^{2(M-k)}}{(M-k)^2N!k!}
{}_4F_3(M\!-\!k,M\!-\!k,M\!+\!1,1;
M\!-\!k\!+\!1,M\!-\!k\!+\!1,N\!+\!1;\varrho_i^2),
\end{split}
\end{equation} where $N=M+\nu$, and
the last line comes from the definition of the generalized
hypergeometric function\cite[pp. 1071,
9.14.1]{IEEE_sw27:RyzhikBook_5th}.

Substitution of (\ref{eq:2ndMoment_MIMO_IMI_HighSNR}),
(\ref{eq:ACF_MIMO_IMI_HighSNR_1}) and (\ref{eq:S(k,nv,varrho)}) into
(\ref{eq:Def_NACF_MIMO_IMI}) results in
(\ref{eq:NACF_HighSNR_MIMO_IMI}). Similarly, with
(\ref{eq:1stMoment_MIMO_IMI_HighSNR}),
(\ref{eq:Variance_MIMO_IMI_HighSNR}),
(\ref{eq:ACF_MIMO_IMI_HighSNR_1}) and (\ref{eq:S(k,nv,varrho)}),
(\ref{eq:Def_Coeff_MIMO_IMI}) reduces to
(\ref{eq:Coeff_HighSNR_MIMO_IMI}).
\section{Proof of Theorem \ref{theo:LCR_AOD_MIMO_IMI_Gaussian_Approx}}
\label{app:LCR_AOD_MIMO_IMI_Gaussian_Approx} To simplify the
notation, we set $X=\mathcal{I}_l$, $Y=\mathcal{I}_{l-1}$, and
$\rho=\rho_{\!\mathcal{I}}(1)$. According to Proposition
\ref{prop:MIMO_IMI_Gaussian_Approx}, we have the PDF of $X$ and the
joint PDF of $X$ and $Y$ as
\begin{equation}\label{eq:PDF_I_l}
p(x)=\frac{1}{\sqrt{2\pi}\sigma_{\!\mathcal{I}}}\exp\left[-\frac{\left(x-\mu_{\!\mathcal{I}}\right)^2}
{2\,\sigma^2_{\!\mathcal{I}}}\right],
\end{equation} and
\begin{equation}\label{eq:jPDF_I_l_I_l-1}
p(x,y)=\frac{\exp\left[-\frac{\left(x-\mu_{\!\mathcal{I}}\right)^2
+\left(y-\mu_{\!\mathcal{I}}\right)^2-2\rho\left(x-\mu_{\!\mathcal{I}}\right)
\left(y-\mu_{\!\mathcal{I}}\right)}
{2\,\sigma^2_{\!\mathcal{I}}\left(1-\rho^2\right)}\right]}
{\sqrt{2\pi\!\left(1-\rho^2\right)}\,\sigma_{\!\mathcal{I}}}.
\end{equation} In what follows, we calculate
$\phi_{\!\mathcal{I}}(I_\mathrm{th})=\int_{I_\mathrm{th}}^\infty
p(x)dx$ and
$\varphi_{\!\mathcal{I}}(I_\mathrm{th})=\int_{I_\mathrm{th}}^\infty
\!\!\int_{I_\mathrm{th}}^\infty p(x,y)dxdy$ for the cases of
$I_\mathrm{th}\geq\mu_{\!\mathcal{I}}$ and
$I_\mathrm{th}<\mu_{\!\mathcal{I}}$.
\subsection{The Case of $I_\mathrm{th}\geq\mu_{\!\mathcal{I}}$}
According to (4.2)\cite{IEEE_sw27:SimonBook04} we obtain
\begin{equation}\label{eq:phi_MIMO_IMI_Gaussian_Approx_Ith>muI}
\phi_{\!\mathcal{I}}(I_\mathrm{th})\operatornamewithlimits{=}
^{I_\mathrm{th}\geq\mu_{\!\mathcal{I}}}
\frac{1}{\pi}\int_0^{\frac{\pi}{2}}
\exp\!\!\left(\!-\frac{\tilde{I}_\mathrm{th}^2}
{2\sin^2\theta}\!\right)\!d\theta,
\end{equation} where $\tilde{I}_\mathrm{th}=\frac{I_\mathrm{th}
-\mu_{\!\mathcal{I}}}{\sigma_{\!\mathcal{I}}}$. Similarly, using
(4.18)\cite{IEEE_sw27:SimonBook04} and the following equality
\begin{equation}
\arctan\left(\sqrt{\frac{1+\rho}{1-\rho}}\right)
=\frac{\pi}{4}+\frac{\arcsin\left(\rho\right)}{2},
\end{equation} we obtain
\begin{equation}\label{eq:varphi_MIMO_IMI_Gaussian_Approx_Ith>muI}
\varphi_{\!\mathcal{I}}(I_\mathrm{th})\operatornamewithlimits{=}
^{I_\mathrm{th}\geq\mu_{\!\mathcal{I}}}
\frac{1}{\pi}\int_0^{\frac{\pi}{4}+\frac{\arcsin\left(\rho\right)}{2}}
\exp\!\!\left(\!-\frac{\tilde{I}_\mathrm{th}^2}
{2\sin^2\theta}\!\right)\!d\theta.
\end{equation} Substitution of (\ref{eq:phi_MIMO_IMI_Gaussian_Approx_Ith>muI})
and (\ref{eq:varphi_MIMO_IMI_Gaussian_Approx_Ith>muI}) into
(\ref{eq:Def_LCR_Up_DT}) results in
(\ref{eq:LCR_MIMO_IMI_Gaussian_Approx}). Moreover,
$F_{\!\mathcal{I}}(I_\mathrm{th})=\int_{-\infty}^{I_\mathrm{th}}
p(x)dx=1-Q\!\left(\tilde{I}_\mathrm{th}\right)$. By plugging
$F_{\!\mathcal{I}}(I_\mathrm{th})$ and
(\ref{eq:LCR_MIMO_IMI_Gaussian_Approx}) into
(\ref{eq:AFD_Eigen-Channel}) we obtain
(\ref{eq:AOD_MIMO_IMI_Gaussian_Approx}).
\subsection{The Case of $I_\mathrm{th}<\mu_{\!\mathcal{I}}$}
For this case, using the results in
(\ref{eq:phi_MIMO_IMI_Gaussian_Approx_Ith>muI}) and
(\ref{eq:varphi_MIMO_IMI_Gaussian_Approx_Ith>muI}) and the symmetry
of the Gaussian PDF, i.e., the integral equality
$\int_{-\infty}^a\frac{1}{\sqrt{2\pi}}
e^{-\frac{t^2}{2}}dt{\displaystyle\operatornamewithlimits{=}^{a\leq0}}\!
\int_{-a}^\infty\frac{1}{\sqrt{2\pi}} e^{-\frac{t^2}{2}}dt$, it is
straightforward to obtain
\begin{equation}\label{eq:phi_MIMO_IMI_Gaussian_Approx_Ith<muI}
\phi_{\!\mathcal{I}}(I_\mathrm{th})\operatornamewithlimits{=}
^{I_\mathrm{th}<\mu_{\!\mathcal{I}}}1-\frac{1}{\pi}\int_0^{\frac{\pi}{2}}
\exp\!\!\left(\!-\frac{\tilde{I}_\mathrm{th}^2}
{2\sin^2\theta}\!\right)\!d\theta,
\end{equation} and
\begin{equation}\label{eq:varphi_MIMO_IMI_Gaussian_Approx_Ith<muI}
\varphi_{\!\mathcal{I}}(I_\mathrm{th})\operatornamewithlimits{=}
^{I_\mathrm{th}<\mu_{\!\mathcal{I}}}1-\frac{2}{\pi}\int_0^{\frac{\pi}{2}}
\exp\!\!\left(\!-\frac{\tilde{I}_\mathrm{th}^2}
{2\sin^2\theta}\!\right)\!d\theta+
\frac{1}{\pi}\int_0^{\frac{\pi}{4}+\frac{\arcsin\left(\rho\right)}{2}}
\exp\!\!\left(\!-\frac{\tilde{I}_\mathrm{th}^2}
{2\sin^2\theta}\!\right)\!d\theta.
\end{equation}
We obtain (\ref{eq:LCR_MIMO_IMI_Gaussian_Approx}) by substituting
(\ref{eq:phi_MIMO_IMI_Gaussian_Approx_Ith<muI}) and
(\ref{eq:varphi_MIMO_IMI_Gaussian_Approx_Ith<muI}) into
(\ref{eq:Def_LCR_Up_DT}). Similarly, we get
(\ref{eq:AOD_MIMO_IMI_Gaussian_Approx}) easily by plugging
$F_{\!\mathcal{I}}(I_\mathrm{th})$ and
(\ref{eq:LCR_MIMO_IMI_Gaussian_Approx}) into
(\ref{eq:AFD_Eigen-Channel}).
\bibliographystyle{IEEEtran}
\bibliography{IEEEabrv,IEEE_sw27}
\begin{table}[htbp]
\renewcommand{\arraystretch}{1.4}
\begin{center}
\begin{threeparttable}
\caption{Taylor Expansion of (\ref{eq:Coeff_HighSNR_MIMO_IMI}) and
the Maximum Difference between (\ref{eq:Coeff_LowSNR_MIMO_IMI}) and
(\ref{eq:Coeff_HighSNR_MIMO_IMI}) for Different $(M, N)$'s}
\label{tab:Taylor_rho_I}
\begin{tabular}{|c||c|c|}
  \hline$(M, N)$&$\text{Taylor Series of}\ (\ref{eq:Coeff_HighSNR_MIMO_IMI})$&
  $\max_{0\leq\varrho_i\leq1}|(\ref{eq:Coeff_LowSNR_MIMO_IMI})-
(\ref{eq:Coeff_HighSNR_MIMO_IMI})|\tnote{\ddag}$ \\
  \hline
  \hline$(1,1)$&$0.608\varrho_i^2+0.152\varrho_i^4+\mathcal{O}(\varrho_i^6)$ &
  $0.160$ \\
  \hline$(2,2)$&$0.437\varrho_i^2+0.218\varrho_i^4+\mathcal{O}(\varrho_i^6)$ &
  $0.230$ \\
  \hline$(3,3)$&$0.372\varrho_i^2+0.186\varrho_i^4+\mathcal{O}(\varrho_i^6)$
  & $0.274$ \\
  \hline$(4,4)$&$0.337\varrho_i^2+0.168\varrho_i^4+\mathcal{O}(\varrho_i^6)$
  & $0.304$ \\
  \hline$(4,8)$&$0.725\varrho_i^2+0.178\varrho_i^4+\mathcal{O}(\varrho_i^6)$ &
  $0.085$ \\
  \hline$\vdots$ & $\vdots$ & $\vdots$ \\
  \hline$(4,12)$&$0.824\varrho_i^2+0.135\varrho_i^4+\mathcal{O}(\varrho_i^6)$
  & $0.050$ \\
  \hline$\vdots$ & $\vdots$ & $\vdots$ \\
  \hline$(4,16)$&$0.870\varrho_i^2+0.107\varrho_i^4+\mathcal{O}(\varrho_i^6)$
  & $0.036$ \\
  \hline
\end{tabular}
\begin{tablenotes}
\item[\ddag] The maximum difference is calculated via the function
\texttt{FindMaximum} in Mathematica$^\circledR$.
\end{tablenotes}
\end{threeparttable}
\end{center}\vspace{-8mm}
\end{table}

\end{document}